\documentclass[aps, reprint, nofootinbib, longbibliography, superscriptaddress]{revtex4-1}

\usepackage{amsmath, amssymb, amsthm}
\usepackage{bm,bbm}

\usepackage{hyperref}
\usepackage[utf8]{inputenc}

\usepackage{soul}
\usepackage{url}
\usepackage{mathtools}
\usepackage{braket}
\usepackage{amsmath}
\usepackage{dsfont}
\usepackage[english]{babel}
\usepackage{graphicx} 
\usepackage{tikz}
\usetikzlibrary{quantikz}

\usepackage{hyperref}
\usepackage{natbib} 
\usepackage{physics}

\usepackage{listings}
\usepackage{xcolor}
\usepackage{amssymb}

\newcommand{\LDIG}{
Lighthouse Disruptive Innovation Group, LLC
7 Broadway Terrace, Apt 1
Cambridge MA 02139
Middlesex County, Massachusetts (USA)
}

\newcommand{\DSD}{
Smart Society Research Group -
La Salle - Universitat Ramon Llull
Carrer de Sant Joan de La Salle, 42
08022 Barcelona (Spain)
}

\newcommand{\MIT}{
MIT Media Lab - City Science Group, Cambridge, USA
}

\usepackage{babel}

\graphicspath{{./images/}}

\begin{document}
\title{Fourier series weight in quantum machine learning}

\author{Parfait Atchade-Adelomou}
\affiliation{\MIT}
\email{parfait@mit.edu}
\affiliation{\DSD}
\email{parfait.atchade@salle.url.edu}
\affiliation{\LDIG}
\email{parfait.atchade@lighthouse-dig.com}

\author{Kent Larson}
\affiliation{\MIT}
\email{kll@media.mit.com}

\date{Feb 2023}

\begin{abstract}
In this work, we aim to confirm the impact of the Fourier series on the quantum machine learning model. We will propose models, tests, and demonstrations to achieve this objective. We designed a quantum machine learning leveraged on the Hamiltonian encoding. With a subtle change, we performed the trigonometric interpolation, binary and multiclass classifier, and a quantum signal processing application. We also proposed a block diagram of determining approximately the Fourier coefficient based on quantum machine learning. We performed and tested all the proposed models using the Pennylane framework.

\end{abstract}

\maketitle

\section{Introduction}\label{sec:intro}

The Fourier series, adept at decomposing complex functions into simpler trigonometric components, aligns seamlessly with quantum computing's intrinsic properties, such as superposition and interference. This synergy results in a more effective and precise representation of quantum information, significantly enhancing data processing, analysis, and exploring periodic patterns within quantum data. This work delves into the profound advantages of Fourier series applications in Quantum Machine Learning (QML), contrasting their unique alignment with quantum computing against conventional methodologies.

The Fourier series is a mathematical tool that allows us to model any arbitrary periodic signal with a combination of sines and cosines. Its main advantage is that more signal information is needed during the transformation from one domain to another. Indeed, this series does not exist for all signals (Dirichlet conditions\cite{chaudhuri2022navier}); however, in various fields and sectors, the Fourier series is the tool used to transform a signal from the time domain to the frequency domain, breaking it down into harmonically related sinusoidal functions.
In quantum computing and specifically in the branch of \textit{quantum machine learning} (QML), a quantum model is described by a parametric function $f(x,\theta)$ subject to some independent variables $x$ that could be our input data and some parameters $\theta$ that help our function to attempt to generalize itself across the input data. Taking this into account and knowing the tremendous impact that the Fourier series has on signal processing, it is therefore of great interest to analyze and experiment to see how the Fourier series impacts the quantum models thus, if it could help us in quantum trigonometric interpolation techniques, regression models, etc.
In this article, we will highlight the importance of the Fourier series in solving real problems with quantum machine learning. We will also propose a generic quantum circuit that, with little change, can help us solve classification problems, interpolation for banking problems, and signal processing, among others.
Complementing this, our method incorporates classical preprocessing techniques to optimize the interplay between data and quantum algorithms. By normalizing and adapting data, we ensure coherent and quantum-compatible inputs, thereby maximizing the efficiency of information encoding within quantum circuits. This holistic approach not only elevates the performance of quantum algorithms but also fosters innovation in the rapidly evolving domain of QML, highlighting the integral role and potential of the Fourier series at the vanguard of this field.

The document is organized as follows. Section \eqref{sec:motivation} presents our primary motivation behind this work. Section \eqref{sec:relatedwork} shows previous work on Fourier series-based quantum machine learning and the simulation of the dynamical behavior of a system. Section \eqref{sec:QML} will present the quantum machine learning framework and its connection to the Hamiltonian simulation. In section \eqref{sec:IMPLEMENT}, we propose the scenarios and the models we will implement to highlight the Fourier Series on quantum machine learning applications. Section \eqref{sec:FS} analyses our model, taking into account some steps from the quantum model's universality. Section \eqref{sec:resultados} presents the obtained results. Section \eqref{sec:discussions} discusses practically relevant results and their implications, and finally, Section \eqref{sec:conclusions} concludes the work carried out, and we open ourselves to some future lines of the proposed model.

\section{Motivation}\label{sec:motivation}
Our primary motivation is to understand the QML model as a Fourier series to join the area of digital signal processing with quantum computing, both analytically and statistically, which is part of quantum machine learning.

\section{Related Work}\label{sec:relatedwork}

In quantum computing, one of the most demanding problems is the simulation of the Hamiltonian. \textit{Hamiltonian simulation} is a problem that requires algorithms that efficiently implement a quantum state's evolution. \textit{Richard Feynman} proposed the Hamiltonian simulation problem in 1982, where he proposed a quantum computer as a possible solution since the simulation of general Hamiltonians seems to grow exponentially concerning the size of the system. Nowadays, this problem is reflected in almost all the systems under study, where we want to analyze the physical design and dynamics within a computational model.

The Hamiltonian simulation problem is defined by the \textit{Schrödinger equation}, where it gives a Hermitian matrix $H(2^n \times 2^n)$ that acts on $n$ qubits in a time $t$ and a maximum simulation error $(\epsilon)$, whose goal is to find an algorithm that approximates an operator $U$ such that $||U- (e^{-iHt})|| \leq \epsilon$. Where $e^{-iHt}$ is the ideal evolution.

The quantum simulation of the dynamical behavior of a system is usually executed in polynomial time $P$, $BQP$, etc. But its Hamiltonian matrix grows exponentially $2^n \times 2^n$ relating to the $n$ qubits of the systems under study. Thus, techniques are sought to tackle the best solution assuming an error. There are two strategies, \textit{Julius Caesar} (divide and conquer) and the \textit{quantum walk algorithm}\cite{ambainis2007quantum, schuld2018supervised}. A helper is a \textit{Local Hamiltonian} for some specific Hamiltonians. A \textit{k-local Hamiltonian} \cite{kempe2006complexity} is a Hermitian matrix acting on $n$ qubits that can be represented as the sum $m$ of the Hamiltonian terms acting on each of the qubits at most: $H = \sum_i H_i$. This also gives rise to \textit{d-sparse} \cite{childs2010simulating}. A Hamiltonian is said to be \textit{d-sparse} (on a fixed basis) if it has at most $d$ nonzero entries in any row or column.

From the \textit{Divide and conquer}, the \textit{first step} breaks down the Hamiltonian into a sum of small and straightforward Hamiltonians, and the \textit{second step} has the objective of recombining the sums of small and straightforward Hamiltonians. To do this, there are three great techniques.
\begin{enumerate}
    \item The first technique is one of the mature approaches, which is $e^{-i( A+B)t}$ where $A$ and $B$ are the small and straightforward Hamiltonians in an approximation of ($e^{-i At/r}e^{-iBt/r})^r$ for large a real value $r$ \cite{hatano2005finding}.
    \item The second technique combines a \textit{Linear Combination of Unitaries (LCU)}\cite{meister2022tailoring, wang2021fast,EVAPG} and the \textit{Oblivious Amplitude Amplification (OAA)} \cite{de1983applications, guerreschi2019repeat, OAA2}.
    \item The latest newest technique is the \textit{Quantum Signal Processing} (QSP) \cite{eldar2002quantum, fourierBasedQSP}.
\end{enumerate}

In contrast, computing a system's ground state energy optimizes a given system's global property. Usually, this problem is very demanding (NP-Hard). The most known problem is calculating the minimum expectation value of a quantum circuit $\langle \phi|H|\phi\rangle$ overall $|\phi\rangle$. 

There are two ways to solve this problem using the \textit{Top-Down} philosophy and statistical methods. On the one hand, the \textit{Top-Down}  is where analytical techniques are applied. Conversely, statistical methods (like QML) are sought through the \textit{variational principle}, where a model approximates the solution. This paper will focus on the second case and highlight the relationship between the two resolution approaches.

In this research, we build upon the foundational work \cite{du2020expressive} and \cite{Schuld_2021}, who have previously explored the representation of quantum models as Fourier series. While several studies \cite{perez2020data, killoran2019continuous, sim2019expressibility,chen2021universal, du2020expressive, biamonte2021universal} have delved into techniques pertinent to our chosen approach, our work uniquely extends these concepts. The existing literature, notably \cite{Schuld_2021}, offers an insightful framework for establishing foundational intuitions in Quantum Signal Processing (QSP) and Quantum Machine Learning (QML). These studies present intriguing practical approaches, yet they often grapple with the inherent complexities of the subject matter, primarily focusing on the theoretical advantages of data processing.
Our article aims to further this exploration in three critical ways. First, we seek to substantiate the natural propensity of quantum models to learn periodic patterns within data. Second, we explore the Fourier series representation in the context of time series learning and signal processing tasks, asserting its essential role in trigonometric interpolation for QML applications. Third, we experiment with and propose a quantum circuit design similar to a neural network, adaptable for interpolation and classification tasks.
Advancing beyond the theoretical insights of \cite{du2020expressive, Schuld_2021}, our research significantly contributes to the practical application of these theories. We offer a comprehensive methodology for the pragmatic computation of Fourier coefficients, tailoring our approach to specific use cases in classification and trigonometric interpolation. Our work bridges the gap between abstract mathematical theories and real-world computational implementations in QML by harmonizing theoretical rigor and practical applicability. It sets a new standard in the field. We aim to showcase the efficacy and flexibility of our method, offering robust solutions to the intricate challenges of quantum data processing and thereby establishing a new benchmark in the realm of Quantum Machine Learning.

\section{Quantum Machine Learning}\label{sec:QML}
Quantum machine learning (QML) \cite{schuld2015introduction, biamonte2017quantum, schuld2018supervised} explores the interplay and takes advantage of quantum computing and machine learning ideas and techniques.

Therefore, quantum machine learning is a hybrid system involving both classical and quantum processing, where computationally complex subroutines are given to quantum devices. QML tries to take advantage of the classical machine learning does best and what it costs, such as distance calculation (inner product), passing it onto a quantum computer that can compute it natively in the Hilbert vector space. 
In this era of extensive classical data and few qubits, the most common use is to design machine learning algorithms for classical data analysis running on a quantum computer, i.e., quantum-enhanced machine learning \cite{dunjko2016quantum, atchade2020using, atchade2022quantum, adelomou2022quantum,Ket.G,schuld2018supervised,consulpacareu2023quantum, atchadeadelomou2021quantum}. 
The usual supervised learning processes within quantum machine learning can be defined as follows:

\begin{itemize}
    \item The \textit{quantum feature map}: It is the data preparation. In the literature, this stage is recognized as State preparation.
    \item The \textit{quantum model}: It is the model creation. In the literature, it is recognized as unitary evolution.
    \item \textit{The classical error computation }: It is the stage of computing the error where the model best approximates the input set; in machine learning, this stage is known as a prediction.
    \item \textit{Observable}: Normally, this stage is included in the error computation. The observable manifest as linear operators on a Hilbert space represents quantum states' state space. The eigenvalues of observable are real numbers corresponding to possible values; the dynamical variable defined by the observable can be measured. We use Pauli's operators as observable. Nevertheless, we can use some linear combination \eqref{eq:obser_Pauli} of these operators or some approximation \eqref{eq:low_weigh_appr} from them to deal with specific measurements.
\end{itemize}

\subsection{Feature Map}
This article deals exclusively with classical data; we receive classical data, and a quantum computer processes it and returns the outcome after the measurement. There are other (three additional) approaches and combinations to process classical or quantum data with hybrid computation \cite{schuld2018supervised}.
In quantum machine learning techniques, some fundamental operations called embedding are performed to treat classical data. 
The embedding process is summarized as classical encoding inputs into quantum states. 
In the literature, several techniques and their relative associated computational costs are found in these references \cite{havlivcek2019supervised,schuld2018supervised, Schuld_2021,schuld2019quantum,perez2020data}. 
We call the \textit{Feature Map} the container carrying out said mapping operation (embedding). The embedding encoding can be amplitude, phase, base, or Hamiltonian, which is the crucial process of QML operations  \cite{schuld2018supervised}.

This work follows the data encoding Hamiltonians idea, which is described as follows: For any unitary matrix $A$, there is a real vector $(\alpha,\beta,\gamma,\delta)$  such that $A = \alpha I+\beta X + \delta Y + \gamma Z$. For large enough $r$, the following equation holds $e^{-i(H_A+H_B)t} \approx (e^{-i H_A t/r}e^{-i H_B t/r})^r$ using the \textit{Trotter Suzuki} formula \cite{de1983applications} which implies that if $r$ scales as $m^2t^2/\epsilon$, the error can be at most $\epsilon$ for any $\epsilon > 0$.

More accurate approximations can be generated by constructing a sequence of exponential operators such that the error terms cancel. The most straightforward formula, which is the second-order \textit{Trotter Suzuki} formula \cite{wittek2013second}, takes the form as follows:
\begin{equation}
    U_2(t) :=\left(\prod_{j=1}^m e^{(-iH_j\frac{t}{2r})}\prod_{j=1}^m e^{(-iH_j\frac{t}{2r})}\right)^r
    \label{eq:trottersuzuki},
\end{equation}
where the error can be given to less than $\epsilon$ for any $\epsilon >0$ when choosing $r$ to scale as $m^{3/2}t^{3/2}/\sqrt{\epsilon}$.

One of the alternative ways to encode the data is by amplitude into a quantum state given by complex continuous data represented as complex values $\boldsymbol{A} \in \mathbb{C}^{2^{N}}$:
\begin{equation}
    \boldsymbol{A}=(a_0, \ldots, a_{2^N-1})\mapsto \sum_{k=0}^{2^N-1}a_{k}|k\rangle
    \label{eq:AmplitudEmbedding},
\end{equation}
with:
\begin{equation}
\lVert\boldsymbol{A}\rVert=1.
    \label{eq:normalize}
\end{equation}

\begin{figure}[htbp]
	\centering
		\includegraphics[width=0.45\textwidth]{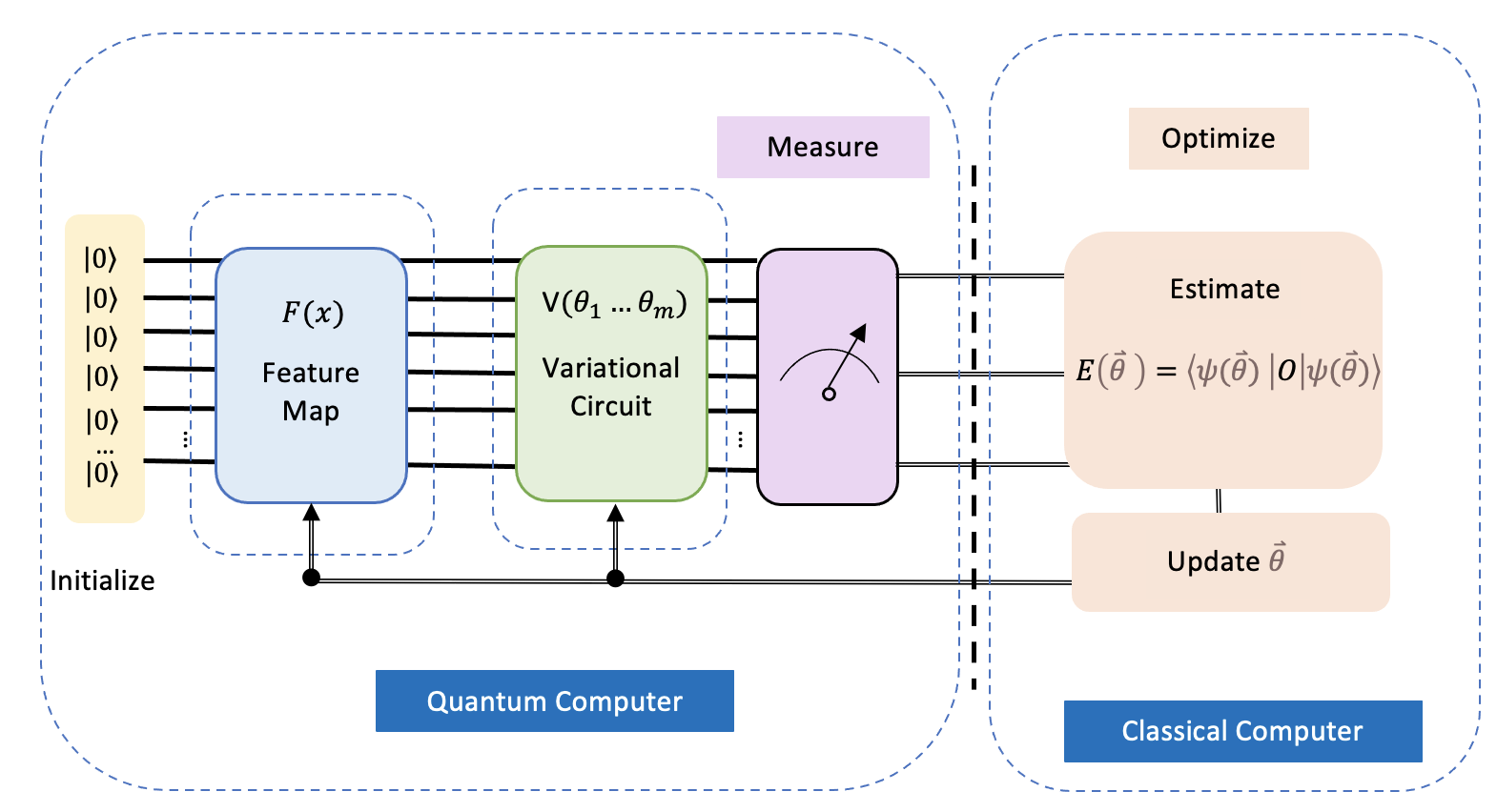}
		\caption{This figure is the standard quantum machine learning model in the literature with only one embedding block and an \textit{Ansatz} as a parameterized quantum circuit with $m$ parameters. This model has a limitation: in the worst case, when the input data is not very well coded, for many parameters $m$ we add to our parametric function/circuit, we cannot find the best model that generalizes the input data.
        There have been various design proposals for how a quantum circuit should be \cite{lloyd2020quantum, suzuki2020analysis, schuld2019quantum}.}
		\label{fig:Std_qModel}
\end{figure}

In the typical design of the quantum model (see Figure \eqref{fig:Std_qModel}), the feature map is usually seen fixed as the only block at the beginning of the quantum circuit and without repetition. Said design/strategy does not help the quantum model generalize the parameterized function better as a neural network, and our proposed model will strongly consider this fact.

\subsection{The model}
This paper focuses on supervised learning applications to drive this section.
Quantum models are parameterizable univariate or multivariate functions. The quantum gates that help to define the parameterizable functions are $RX$, $RY$, and $RZ$. Considering a quantum computer is a stochastic machine, to have a deterministic quantum model, the output of our model must be an expected value (here, considering the computation basis $p(0)-p(1)$). That is, we must average the stochastic quantum computer's outcome. An example of a quantum model is given as follows: 
\begin{equation}
    f_{\boldsymbol \theta,  \boldsymbol \beta}(x) = \bra{0} U^{\dagger}(\vec{x},\beta, \vec{\theta}) O U(\vec{x}, \beta, \vec{\theta}) \ket{0},
    \label{eq:model}
\end{equation}
Where $O$ is our observable, $U(\vec{x}, \beta, \vec{\theta})$ the parameterized circuit with the input data $\Vec{x}$, $\beta$, the scaling factor and $\vec{\theta}$, the parameterized variable.

\subsection{The error calculation}
The error analysis of a QML is identical to the phases of the classical machine learning process. Error analysis allows us to know how to improve the performance of our quantum model. Equation \eqref{eq:err_qml} helps to analyze the training errors.  
\begin{equation}
q\_err = \sqrt{\left(\frac{1}{n}\right)\sum_{i=1}^{n}(q\_model(\vec{x},\theta_i) - Label_i)^{2}}
   \label{eq:err_qml}
\end{equation}

\subsection{The Observable}
In a quantum circuit, the observable determines the form and the function's degree we use to classify or interpolate.
For example, in the case of linear classification, the observable defines the linear form that the model will try to adapt to the input data better.

In our case, we use the computational basis $Z$. However, we can use the observable, such as above mentioned:
\begin{equation}
O = \sum_{P \in \{I,X,Y,Z\}^{\otimes n}}\alpha_pP,
   \label{eq:obser_Pauli}
\end{equation}
and 
\begin{equation}
O^{\text{(low)}} = \sum_{|P| \leq k}\alpha_pP.
   \label{eq:low_weigh_appr}
\end{equation}

\subsection{Limitations of the Proposed Quantum Machine Learning Framework}
\label{subsec:limitations}

Despite the promising advancements presented in our QML framework, it is imperative to acknowledge and address its inherent limitations. This critical examination provides a balanced perspective of our work and paves the way for future research directions.
The current state of quantum computing hardware \cite{atchadeadelomou2022quantum,gonzalez2023quantum} is at the forefront of our limitations. The qubits' scarcity and susceptibility to errors and decoherence significantly cap the complexity and size of the models we can reliably implement. Furthermore, the scalability of quantum hardware remains a paramount challenge, as increasing the number of qubits often leads to heightened error rates and resource management complexities.
The design and implementation of QML algorithms pose their unique challenges. Still in its nascent stages, Quantum programming requires a deep understanding of determining the barren plateau \cite{mcclean2018barren}. Additionally, the optimal design of quantum algorithms that can fully leverage the theoretical capabilities of quantum computing remains an area of ongoing research.
While these limitations delineate the current boundaries of our framework, they also highlight critical areas for future investigation. Addressing these challenges will refine our existing model and contribute significantly to the broader field of QML. Our ongoing research efforts are thus dedicated to exploring these avenues to achieve more robust, scalable, and accessible quantum computing applications \cite{gonzalez2022gps,atchadeadelomou2023efficient}.

\section{IMPLEMENTATION}\label{sec:IMPLEMENT}
\begin{figure}[htbp]
	\centering
		\includegraphics[width=0.45\textwidth]{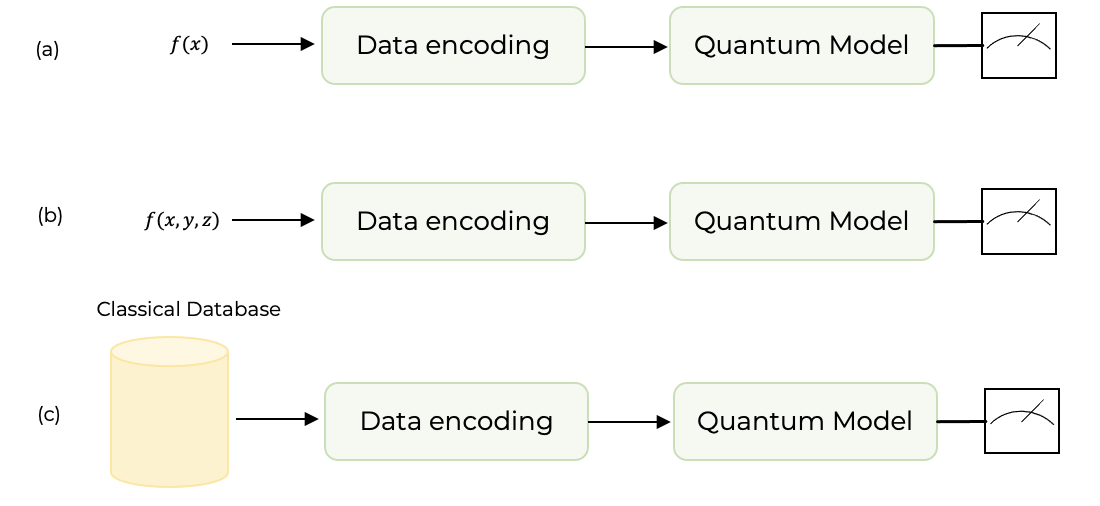}
		\caption{Scenario (a) will be a univariate function. This function will be a \textit{sine}, a \textit{cosine}, a \textit{logarithm}, and a \textit{rectangular pulse}. Scenario (b) is a multivariate function that can be an arithmetic operation on $m$ univariate functions. Scenario (c) is where the input data is regrouped in a classical dataset. Said dataset is the result of a study on classical machine learning models. However, the data can be from any analysis, even if quantum processes after being observed.}
		\label{fig:escenario2}
\end{figure}

The scenarios we follow to validate our experimentation are in figure \eqref{fig:escenario2}. The first scenario is a univariate function. This function will be a \textit{sine}, a \textit{cosine}, a \textit{logarithm}, and a \textit{rectangular pulse}. The second scenario is a multivariate function, an arithmetic operation on $m$ univariate functions. The third scenario is where the input data is regrouped in a classical dataset. Said dataset results from a study on classical machine learning models and statistical events. However, the data can be from any analysis, even if quantum processes after being observed.

\subsection{OUR QUANTUM MODEL} \label{sec:modelTSP}

\begin{figure}[htbp]
	\centering
		\includegraphics[width=0.45\textwidth]{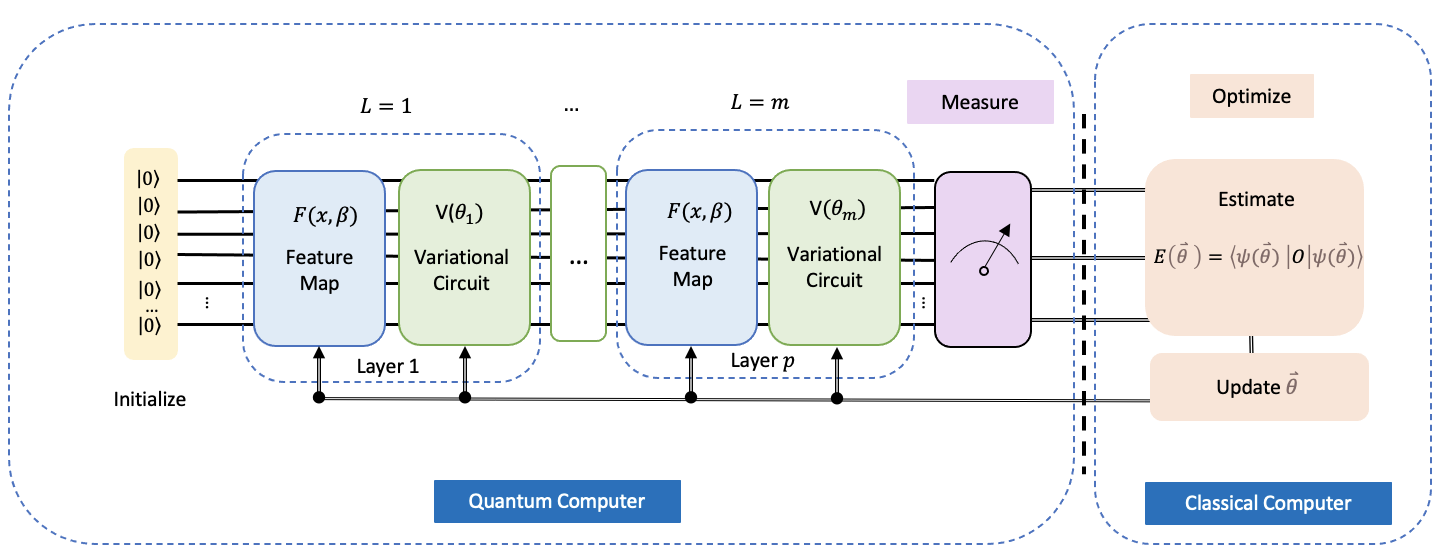}
		\caption{This is the model we use to analyze the weight of the \textit{Fourier series} in quantum machine learning. Depending on some scaling parameters $\beta$ and the input data $\vec{x}$, the Feature Map will be responsible for coding our data. We rely heavily on the fact that the Feature Map ($F(\vec{x},\beta)$) must be variational, that the data's loading is repeated in all the layers, and that the variational circuit ($V(\vec{\theta})$) searches the best function within the space of functions that defines the capacity of the variational circuit ($U(\vec{x},\beta,\vec{\theta})= F(\vec{x},\beta)V(\vec{\theta)}$). With this configuration, we get closer to a quantum neural network that interpolates like a trigonometric function or classifies depending on the problem.}
		\label{fig:qModel}
\end{figure}

Based on the following work \cite{Schuld_2021}, where the authors studied the \textit{expressivity} of the quantum model with a Pauli-rotation and later extended their study in the generality of the quantum model as a Fourier series, we aim to extend said the investigation into a demanding quantum circuit for machine learning. To realize our study, we have defined a \textit{quantum variational Feature Map} together with a \textit{quantum variational circuit} that, taking advantage of its mixture, can behave like a proper quantum neural network and, then verify the Fourier series' weight in those quantum models (see Figure \eqref{fig:qModel}).
With a few subtle changes in the created quantum model, we have performed regression operations, classifications, and trigonometric interpolation.
Let $F(\vec{x},\beta_{i})$ be our feature map and $V(\vec{\theta}_{i})$ be the variational quantum circuit, so the proposed model can be written as follow:
\begin{equation}
    U(\vec{x},\beta, \vec{\theta}) = F(\vec{x},\beta)V(\vec{\theta}),
    \label{eq:proposed_QC}
\end{equation}
where $\vec{x}$ is the input data, $\beta$, the scaling factor, and $\vec{\theta}$ the parameter.

\subsection{Scalability and Adaptability of the Proposed Model}
\label{subsec:scalability_adaptability}
In addressing scalability and adaptability, our model represents a significant advancement in QML. This section elucidates how our proposed framework scales with increasing data complexity and adapts to various quantum computing architectures.
Our model's scalability is a pivotal feature designed to handle the exponential growth of quantum states as system size increases. By leveraging efficient quantum circuit designs and optimizing qubit usage, we ensure that our model remains computationally feasible even as the number of qubits grows. Additionally, we discuss the implementation strategies employed to mitigate issues related to quantum noise and error rates, which typically escalate with larger quantum systems.
Our proposed model's enhanced scalability and adaptability mark a substantial contribution to the field of QML. By addressing these fundamental aspects ( trigonometric interpolation and classifiers), we pave the way for more versatile and powerful quantum computing applications, responding to the burgeoning needs of this rapidly advancing field.

\section{FOURIER SERIES BASED ON QUANTUM CIRCUIT}\label{sec:FS}

In this session, we recover the work \cite{Schuld_2021} on the quantum model's universality and will adapt it to our case.

Based on \cite{Schuld_2021,gil2020input}, where the authors demonstrated the universality of their theorem, considering the multivariate function of equation \eqref{eq:model}. We follow the steps described in said work and confirm that $f(x)$ follows the Fourier series decomposition for one layer. For that, let us recall the equation \eqref{eq:model}, which adapts it to the definition of the proposed feature map $F(\Vec{x},\beta)$ and variation quantum circuit $V(\vec{\theta_i})$ as follow.
\begin{equation}
    f(x) = \bra{0} U^{\dagger}(\vec{x},\vec{\theta},\beta) O U(\vec{x}, \vec{\theta},\beta) \ket{0},
    \label{eq:model_multi}
\end{equation}

where
\begin{equation}
    U(\vec{\theta},\beta,\vec{x}) = V^{(2)}(\vec{\theta}^{(2)})F(\vec{x},\beta) V^{(1)}(\vec{\theta}^{(1)}),
\end{equation}
with $\vec{\theta}^{(1)}, \vec{\theta}^{(2)} \subseteq \vec{\theta}$ and recalling the equation \eqref{eq:proposed_QC}, we can re-write it as follows:
\begin{equation} \label{eq:Feat_multi}
    F(\vec{x},\beta) = e^{-i x_1 \beta H_1}\otimes\ldots\otimes e^{-i x_N \beta H_N }.
\end{equation}

If we make the assumption that $H_p$ is diagonal and its eigenvalues are $\{\lambda_1^{p},\ldots,\lambda_N^{p}\}$, so, $H_p = diag(\lambda_1^{p},\ldots,\lambda_{2^d}^{p})$. Let us define $\Lambda$ as follows:

\begin{equation}
    \Lambda := (\lambda_{j1}^{p},\ldots,\lambda_{jN}^{p}),
\end{equation}


Let $d$ be the number of qubits of our quantum system, let $\Vec{j}$ be $\{j_1,\ldots,j_N\}$ and $\vec{k}$ be $\{k_1,\ldots,k_N\}$ $\in [2^d]^N$, then, let us define $[F(\vec{x},\beta)]$ as follows taking in account the previous definition:
\begin{equation}
    [F(\vec{x},\beta)]_{\vec{j},\vec{j}} := e^{-i\vec{x}\beta\cdot \vec{\Lambda}_{\vec{j}}}.
\end{equation}

Following the same assumption that we will confirm by experimentation on the trainable circuit blocks \eqref{fig:VQA_Fourier_Coeff}, we "drop" the explicit dependence on $\vec{\theta}$, and by absorbing $V^{(1)}$ into the initial state $\ket{\Phi}$, and $V^{(2)}$ into the observable $O$, consider the following model:
\begin{align}\label{eq:multivariate_model}
    f(\vec{x}) = \langle \Phi | F^{\dagger}(\vec{x},\beta) O F(\vec{x},\beta) |\Phi\rangle,
\end{align}
where
\begin{align}\label{eqn:gama_}
    |\Phi\rangle &= \sum_{j_1,\ldots,j_N = 1}^{2^d}\alpha_{j_1,\ldots,j_N}^{\vphantom{*}}|j_1\rangle\otimes\ldots\otimes|j_N\rangle.
\end{align}

Let us note $\alpha^*_{\vec{j}}$ and $\alpha_{\vec{k}}$ as a complex nunmber as follows:
\begin{equation}
    \alpha^*_{\vec{j}} = \alpha^*_{{j_1,\ldots,j_N}},
\end{equation}
\begin{equation}
    \alpha_{\vec{k}} = \alpha_{{k_1,\ldots, k_N}}.
\end{equation}

Then let us re-write \eqref{eqn:gama_} as follows: 
\begin{align}
    |\Phi\rangle :=\sum_{\vec{j}}^{}\alpha{\vec{j}}|\vec{j}\rangle.
\end{align}

Thus, the associated quantum model is defined as follows:
\begin{align}\label{eqn:model_expression}
    f_{\text{Fourier}}(\vec{x}) &= \sum_{\vec{j}}\sum_{\vec{k}} \alpha^*_{\vec{j}}\alpha_{\vec{k}}^{\vphantom{*}}O_{\vec{j},\vec{k}}^{\vphantom{*}}e^{i\vec{x}\beta \cdot(\vec{\Lambda}_{\vec{k}} - \vec{\Lambda}_{\vec{j}})},
\end{align}

where the multi-indices $\vec{j}$ and $\vec{k}$ have $N$ entries that iterate over all $2^d$ basis states of the $d$ qubit subsystems. 
Let us define $\omega_{H}$ as the frequency spectrum of $H$ as $\omega_{H} = \{\Lambda_{\vec{j}}-\Lambda_{\vec{k}}\,|\, \vec{j},\vec{k} \in [2^d]^N\}$.

The equation \eqref{eqn:model_expression}  is, in fact, a partial multivariate Fourier series, with the accessible frequencies entirely determined by the spectra of the coding Hamiltonians ${H_k}$ and the Fourier coefficients determined by the unitary estimates.

The strategy we follow to compute the \textit{Fourier coefficients} of a given function $f(x)$ is measuring the expectation value of the circuit (equation \eqref{eqn:expectedVal}) and classically computing the $c_{n_1,\dots, n_N}$ as an approximation. In our case, the expectation value of our circuit is given by equation \eqref{eqn:model_expression}, and the result of the computation is the form of equation \eqref{eqn:components_ser}.

\begin{align}\label{eqn:expectedVal}
    \langle \hat{O} \rangle & =  \bra{0} U^{\dagger}(\vec{x},\vec{\theta},\beta) \hat{O}U(\vec{x}, \vec{\theta},\beta) \ket{0} \\
    &= \langle \psi(\vec{x},\vec{\theta},\beta) \vert \hat{O} \vert \psi (\vec{x},\vec{\theta},\beta)\rangle.
\end{align}

\begin{align}\label{eqn:components_ser}
f(x) = \sum \limits_{n_1\in \omega_1} \dots \sum \limits_{n_N \in \omega_N} c_{n_1,\dots, n_N} e^{-i x_1 n_1} \dots e^{-i x_N n_N}.
\end{align}

\subsubsection{OUR MODEL AS A TRIGONOMETRIC INTERPOLATION}
From the section \eqref{sec:FS} we can assume that $f(x)$ has Fourier series decomposition. So, $f(x)$ can be written as follows: 
\begin{equation}
    f(x) = c_0 + \sum_{n=0}^{n-1} c_{n}(\theta) e^{i n x},
    \label{eq:appro_trigo_1}
\end{equation}

where the Fourier coefficients $c_{n}$ are given by:
\begin{equation}
    c_n = \frac{1}{\sqrt{2\pi}} \int_{0}^{2\pi}e^{-i w x}f(x) dx,
\end{equation}

We assume that the series converges uniformly to $f(x)$ thus, 
\begin{equation}
    |c_n (w)| \leq \frac{\text{const}}{{|w|^{p+1}+1}}.
    \label{eq:Fourier_decay}
\end{equation}

The Fourier series converges faster for smoother functions (equation \eqref{eq:Fourier_decay}), which is observed by integrating equation \eqref{eq:appro_trigo_1}. By increasing the smoothness one step, the Fourier coefficients decay one step faster as functions of $w$. With $p$, the $pth$ derivative of Fourier coefficients. Which generalizes with Parseval's theorem \cite{sakurai1938extension}.

In the results section, we will have the test benches we did to confirm that trigonometric interpolation can be done with the quantum model. 

In this case, the number of the loading data, the depth of the quantum circuit, the layer ($l$), or repetitions $r$ is proportional to the degree of the function to be interpolated.

\subsubsection{HAMILTONIAN SIMULATION}

As we introduced above, the strategy used to encode the data is the Hamiltonian data encoding defined by equation \eqref{eq:trottersuzuki}, which allows us to deal with quantum signal processing as a strategy to recombine the \textit{Trotter-Suzuki} action on the given Hamiltonian.
To validate our quantum model, we have used this test example: 
The time evolution of the time-independent Hamiltonian $\hat{H}$ is given by $e^{i \hat{H} t}$. Considering that $E_\lambda$ is the energy eigenvalues for eigenstate $|\lambda\rangle$, then

$$ e^{i \hat{H} t} = \sum_\lambda e^{i E_\lambda t} |\lambda\rangle \langle \lambda | =  \sum_\lambda (\cos  E_\lambda t + i \sin E_\lambda t) |\lambda\rangle \langle \lambda |.$$

Thus our aim was simulating $p(x) = \cos(t x)$ and $g(x) = \sin(tx)$ for some instant $t$. We will discuss the results obtained in the results \eqref{sec:resultados} and discussion \eqref{sec:discussions} sessions.






\subsubsection{OUR MODEL AS A CLASSIFIER}
This session will use the quantum model from equation \eqref{eq:model_multi} to achieve the multi-class classifier.

Let $n$ be the number of qubits, let $ \vec {x} $ be a vector of dimension $m$, let $\vec {\theta}$, a matrix of dimension $n \times m $, let $\Vec{\beta}$ the scaler, we can define our model as follows:
\begin{equation}
U (\vec {x}, \vec {\theta}, \beta) = \bigotimes_{i = 0} ^ {n - 1} U_{i} (\vec {x}, \beta_i, \vec {\theta_i}),
    \label{eq:U_teta}
\end{equation}
where:
\begin{equation}
U_{i} (\vec {x} , \beta, \vec {\theta_i}) = F(\vec {x_i}, \beta_i) V(\vec {\theta_i}).
    \label{eq:U_teta_i}
\end{equation}

\begin{figure}[htbp]
	\centering
		\includegraphics[width=0.45\textwidth]{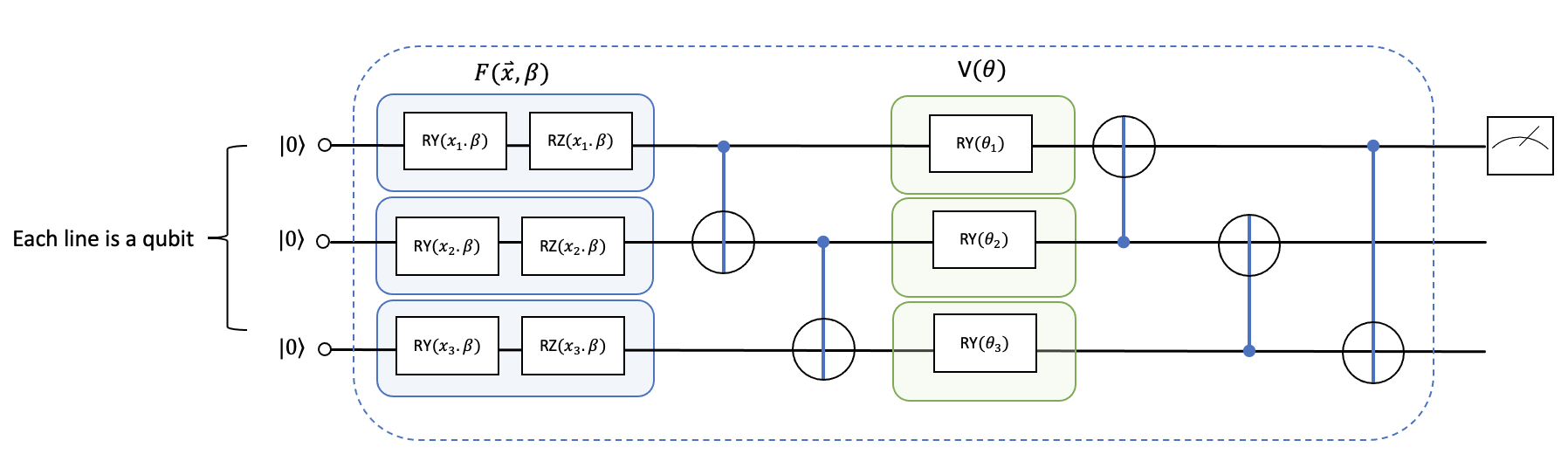}
		\caption{This figure shows a quantum circuit of three qubits with only one repetition (Layer). This circuit implements a trigonometric interpolation (\emph{Fourier series}), where the repeat frequency of the \emph{feature map}, thus of the data, defines the angular frequency of the \emph{Fourier series}: $\sum_{\omega} c_{\omega} e^{i \omega x}$. Our \emph{Feature Map}, which implements our data encoding strategy, determines the frequencies $\omega$, and our \emph{variational quantum circuit} determines the coefficients $c_{\omega}$. Each qubit represents a variable of our multivariable function. The function implements the feature map in blue, with $\beta$, the scaling hyperparameter, with $x$ the input data. Said data is encoded in the $RY$ and $RZ$ parameterized gates. In green is the variational circuit that forms the proposed parameterized function, with $\theta$, our parameter, one parameter per qubit. We have three. The $CNOT$ gates help us interleave the data so we can only read in the first qubit. This allows us to reduce the effect of the barren plateau in the case of a classifier.}
		\label{fig:qCircuit_1rep}
\end{figure}

With $F(\vec {x_i}, \beta_i)$ our feature map function and $V(\vec {\theta_i})$ our variational quantum circuit as it can be seen in Figure \eqref{fig:qCircuit_1rep}.

\begin{figure}[htbp]
	\centering
		\includegraphics[width=0.4\textwidth]{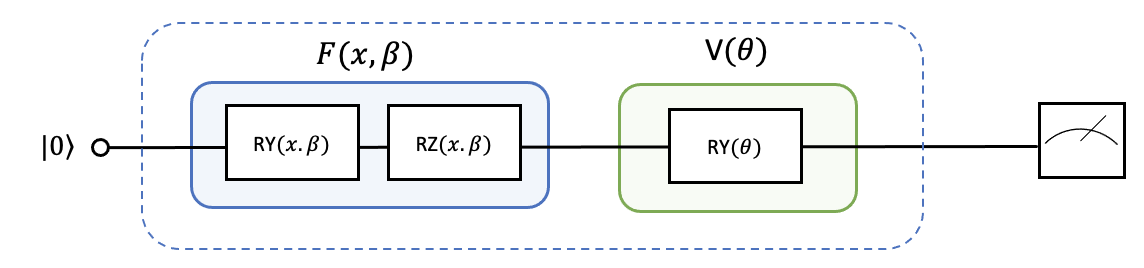}
		\caption{This figure shows a quantum circuit of one qubit with only one repetition (Layer). This circuit implements a trigonometric interpolation (\emph{Fourier series}), where the repeat frequency of the \emph{feature map}, thus of the data, defines the angular frequency of the \emph{Fourier series}: $\sum_{\omega} c_{\omega} e^{i \omega x}$. Our \emph{Feature Map}, which implements our data encoding strategy, determines the frequencies $\omega$, and our \emph{variational quantum circuit} determines the coefficients $c_{\omega}$. In blue, the function that implements the feature map, with $\beta$, the scaling hyperparameter, with $x$ the input data. Said data is encoded in the $RY$ and $RZ$ parameterized gates. In green is the variational circuit that forms the proposed parameterized function, with $\theta$, the parameter, one parameter per qubit. Without repeating the layer and reloading the data $x$, this circuit will only learn a $sine$ or $cosine$ function.}
		\label{fig:qCircuit_1_1rep}
\end{figure}

Let $\ket{\psi(\vec{x})}$ be a functional quantum state and let $f_{\theta_i}(x): \mathbb{R}^{m} \rightarrow \mathbb{C}$ be complex function. Then,

\begin{equation}
\ket{\psi(\vec{x})} = \sum_{i=0}^{2^{n}-1}f_{\theta_i}(\vec{x}) \ket i,
    \label{eq:phi_out}
\end{equation}

\begin{equation}
\sum_{i=0}^{2^{n}-1}  |f_{\theta_i}(\vec{x})|^2 = 1.
    \label{eq:Completeness}
\end{equation}

The circuit $\mathcal{U}(\vec{x}, \vec{\theta}, \beta)$ approximates the state as
\begin{equation}
\ket{\psi(\vec{x})} \sim \mathcal{U}(\vec{x}, \vec{\theta}, \beta) \ket{0}^{\otimes n}, 
    \label{eq:U_hat}
\end{equation}
with
\begin{equation}
\mathcal{U}(\vec{x}, \vec{\theta}, \beta) = \prod_{i=1}^k U(\vec{x}\cdot \beta,\vec{\theta_i}),
    \label{eq:U_hat_1}
\end{equation}
with better results as the number of layers, repetition $k$ increases, and $n$ the number of classes.
$\vec{\theta} = \{ \vec{\theta_i}\}$ is found with classical optimization techniques.
Our Cost function (CF) is calculated by ${\rm distance}(\ket{\psi(\vec{x})}, \mathcal{U}(\vec{x},\beta, \vec{\theta}) \ket{0}^{\otimes n}).$

With all the last defined, the prediction of the quantum model for each $x$ can be defined as the expectation value of the observable $\sigma_z$ concerning the state $\ket{\psi(\vec{x})}$ via our parametrized quantum circuit as follows:

\begin{equation}
    f_{\boldsymbol \theta}(x) = \bra{0} \mathcal{U}^{\dagger}(\vec{x},\vec{\theta},\beta) O \mathcal{U}(\vec{x}, \vec{\theta},\beta) \ket{0},
    \label{eq:model_11}
\end{equation}
where $O$ is observable that can be one of the Pauli matrices $(I, X, Y, Z)$. 

In the case of the binary classifier given by the figure \eqref{fig:qCircuit_1_1rep}, we measure the expected value of the proposed quantum circuit, and after the measurement, we make the sign of the output.

Two possible scenarios exist for the multiclass classifier given by figure \eqref{fig:qCircuit_1rep}. First, we generalize the binary classifier, or second, where we work with the probabilities so that, knowing the highest probability, the class will be known.
Although the problem requires many more qubits due to the input data size, in this case, we use gates like $CNOT$, $CRY$, etc., to entangle the qubits and only read, for example, at qubit 0. Although the number of classes grows logarithmically, the effect of the \textit{Barren Plateaus} \cite{holmes2022connecting, mcclean2018barren}  is less.

\section{Results} \label{sec:resultados}
Considering the proposed scenarios, this session will present the results obtained during our experimentation.

\subsubsection{TRIGONOMETRIC INTERPOLATION}
\begin{figure}[]
	\centering
        \includegraphics[width=0.2\textwidth]{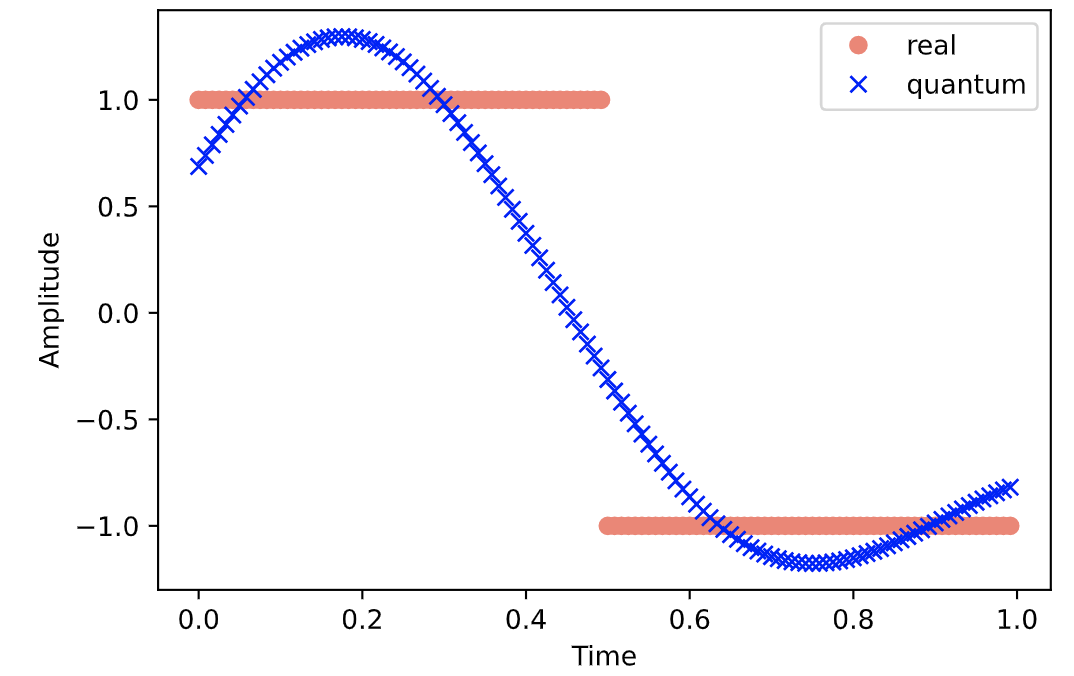}
        \includegraphics[width=0.2\textwidth]{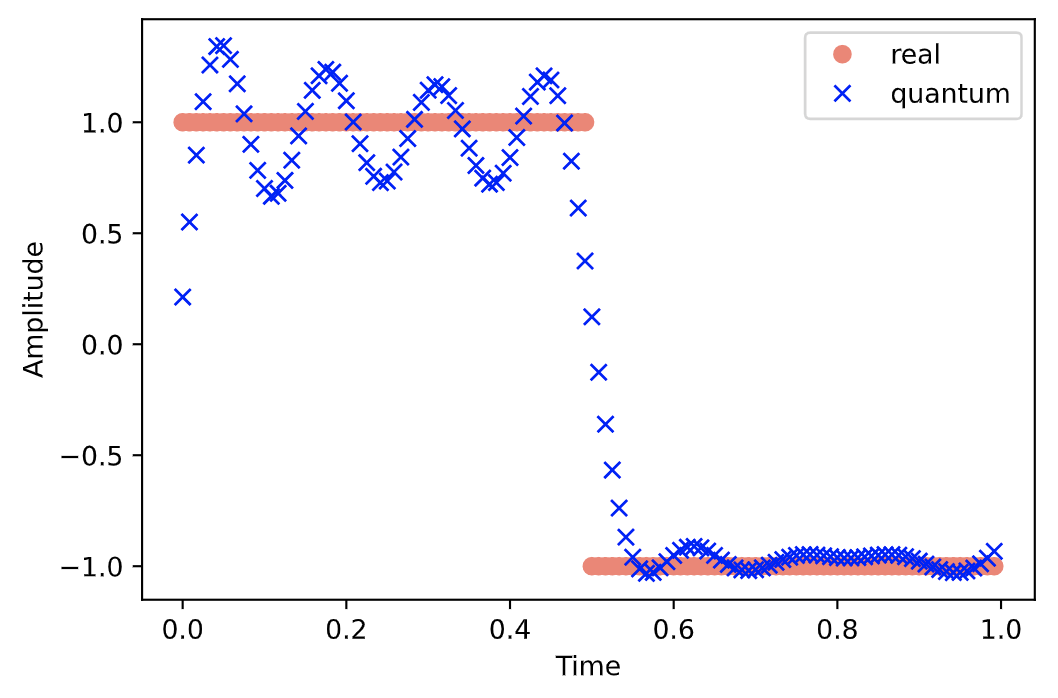}
		\includegraphics[width=0.2\textwidth]{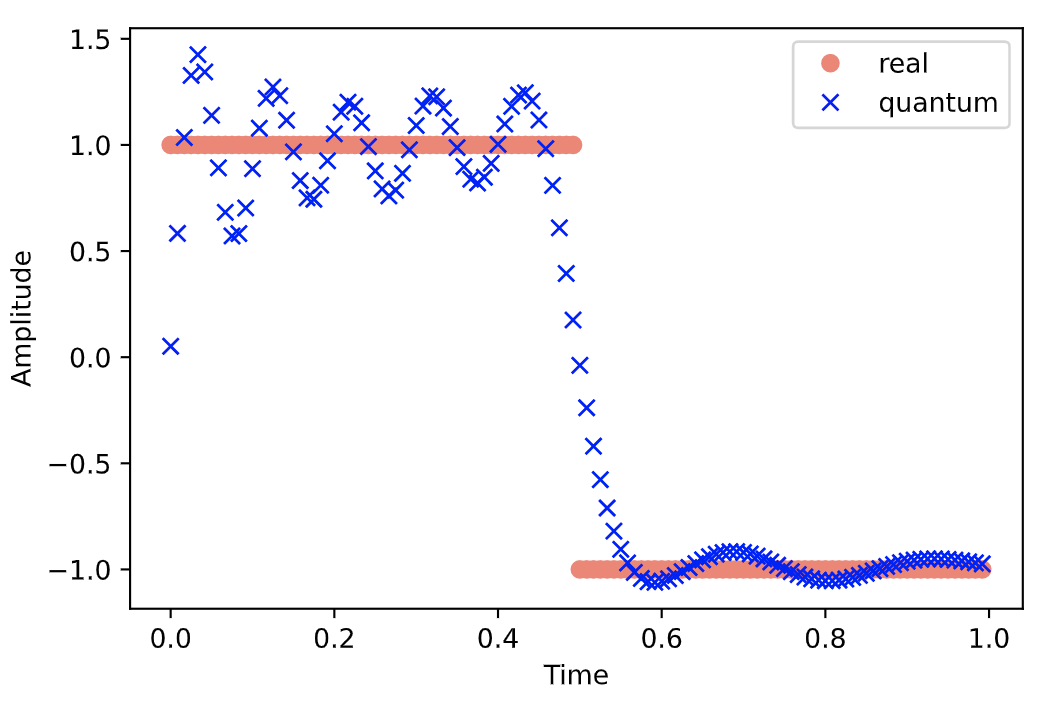}
        \includegraphics[width=0.2\textwidth]{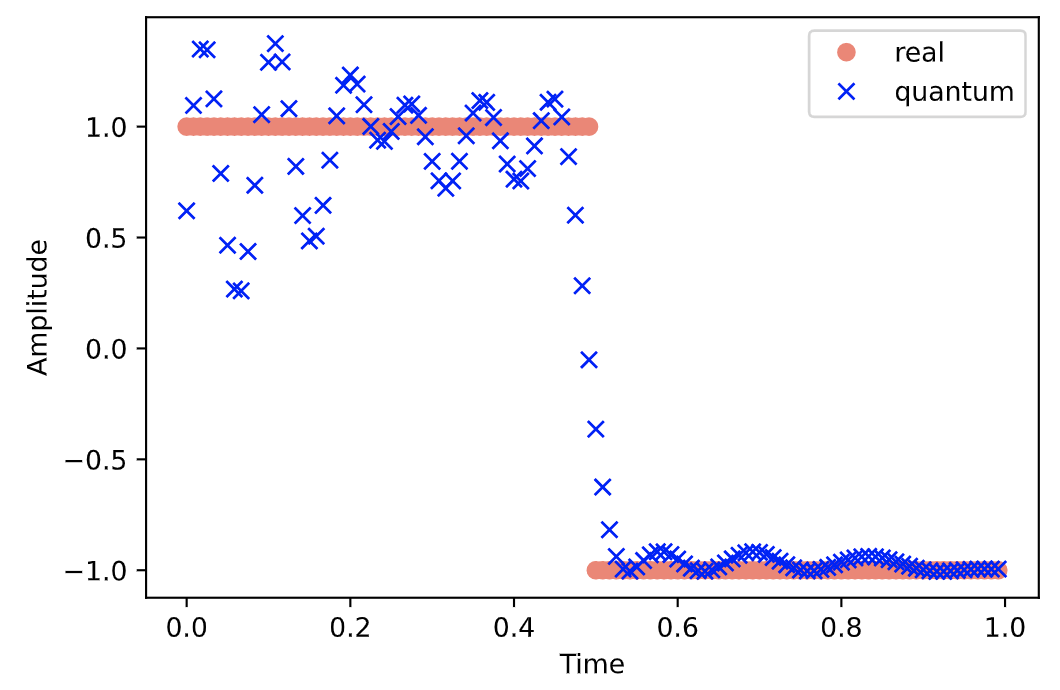}
		\caption{These graphs show the experiment of four tests with the square signal of $20Hz$ with the different number of layers ($L$) and sample ($n$) where the behavior of the trigonometric model on the given input and the decomposition of harmonic frequencies and the \textit{Gibbs phenomenon} can be observed. With more high-frequency components, the reconstruction of the squared signal is achieved. In the first image on the top left, we observe the effect of the Fourier series approximation with a lower-order polynomial. In this case, with the number of the layers $L=3$ and the number of the sample $n= 20$. The second image above to the right is the same scenario as the previous one but with $n=100$. In the third image from the bottom left, we can observe the same input signal of $20Hz$, this time with $L= 20$ and the number of points $n=100$. The fourth image from the bottom right shows the same $20Hz$ input signal, with $L=30$ and $n=100$. We have a scaling parameter $\beta$ to better interpolate with high-frequency harmonics in all these cases.}
		\label{fig:resTrigoSquare}
\end{figure}

\begin{figure}[]
	\centering
		\includegraphics[width=0.2\textwidth]{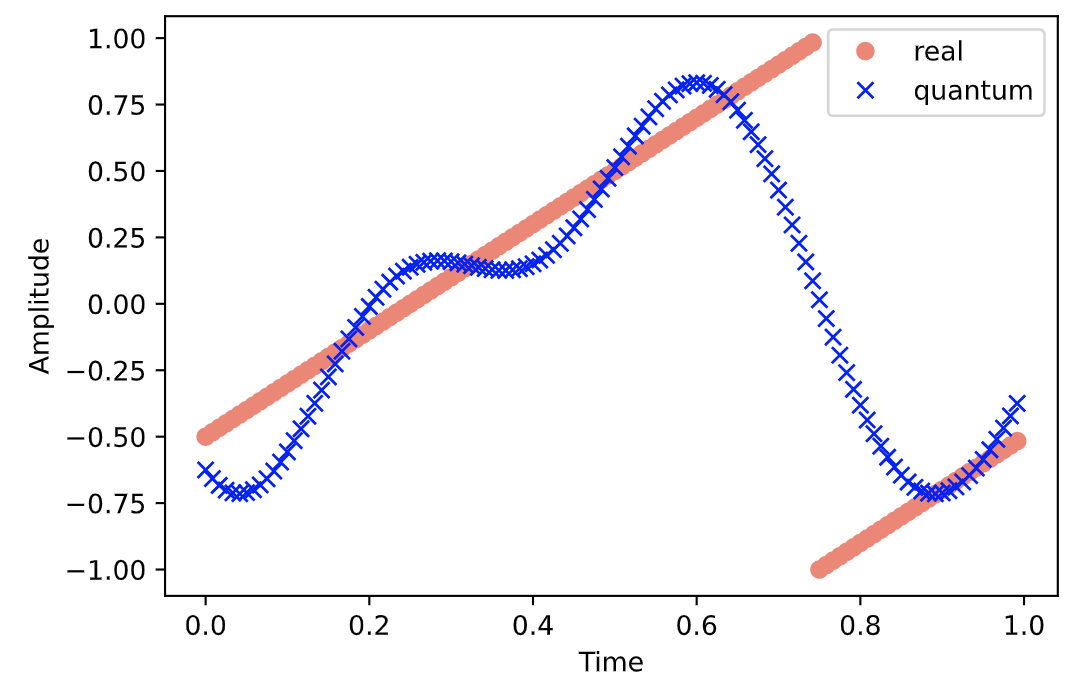}
        \includegraphics[width=0.2\textwidth]{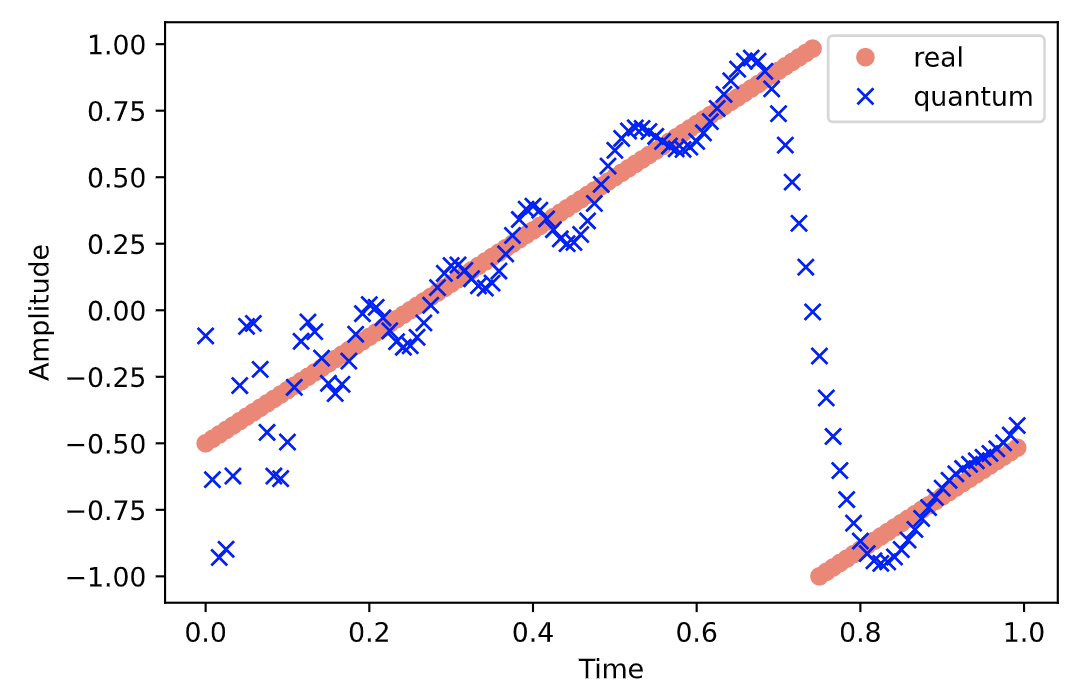}
        \includegraphics[width=0.2\textwidth]{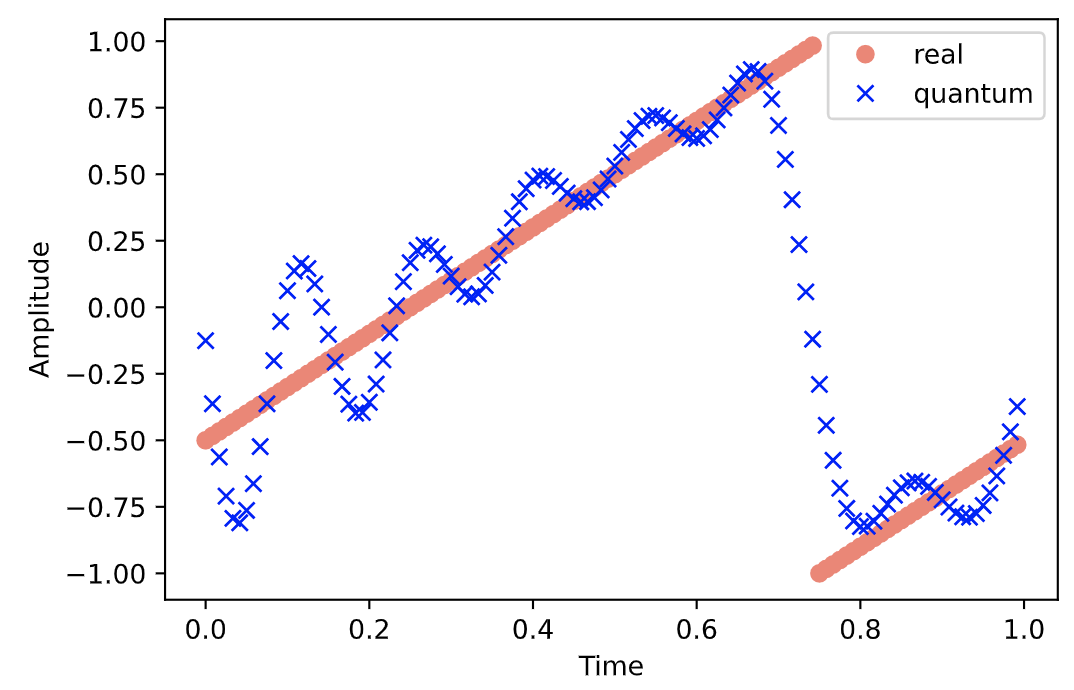}
        \includegraphics[width=0.2\textwidth]{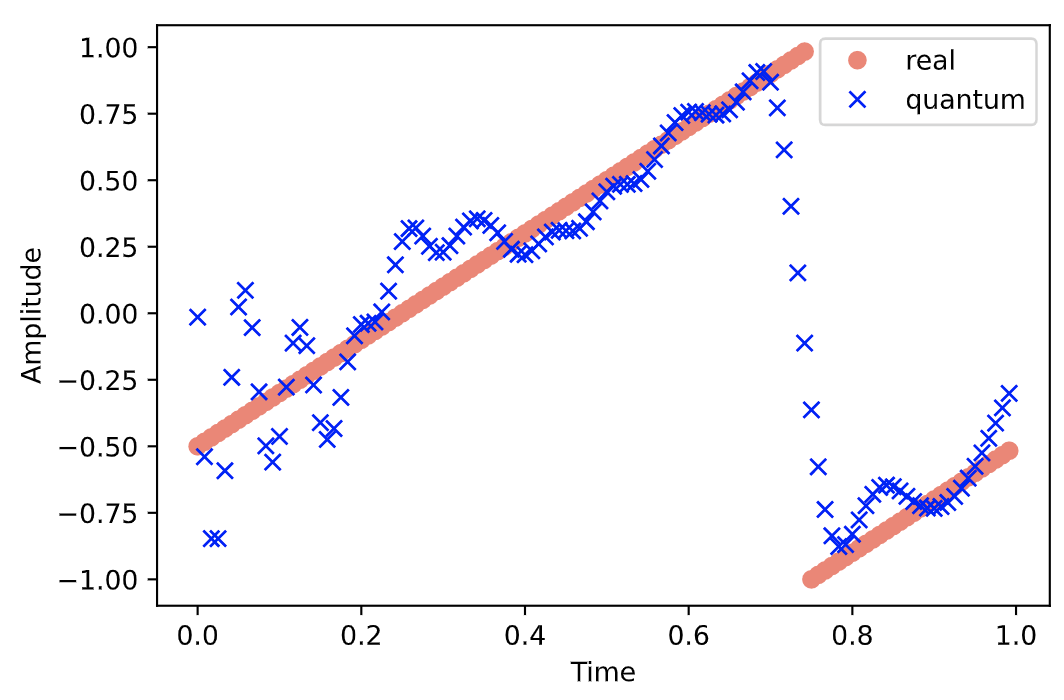}
		\caption{These graphs show the experiment of four tests with the sawtooth signal of $20Hz$ with the different number of layers ($L$) and sample ($n$) where the behavior of the trigonometric model on the given input and the decomposition of harmonic frequencies and the \textit{Gibbs phenomenon} can be observed. With more high-frequency components, the reconstruction of the squared signal is achieved. In the first image on the top left, we observe the effect of the Fourier series approximation with a lower-order polynomial. In this case, with the number of the layers $L=3$ and the number of the sample $n= 20$. The second image above to the right is the same scenario as the previous one but with $n=100$. In the third image from the bottom left, we can observe the same input signal of $20Hz$, this time with $L= 20$ and the number of points $n=100$. The fourth image from the bottom right shows the same $20Hz$ input signal, with $L=30$ and $n=100$. We have a scaling parameter $\beta$ to better interpolate with high-frequency harmonics in all these cases.}
		\label{fig:resTrigoSquare.}
		\label{fig:resTrigoSaw}
\end{figure}

\begin{figure}[]
	\centering
        \includegraphics[width=0.2\textwidth]{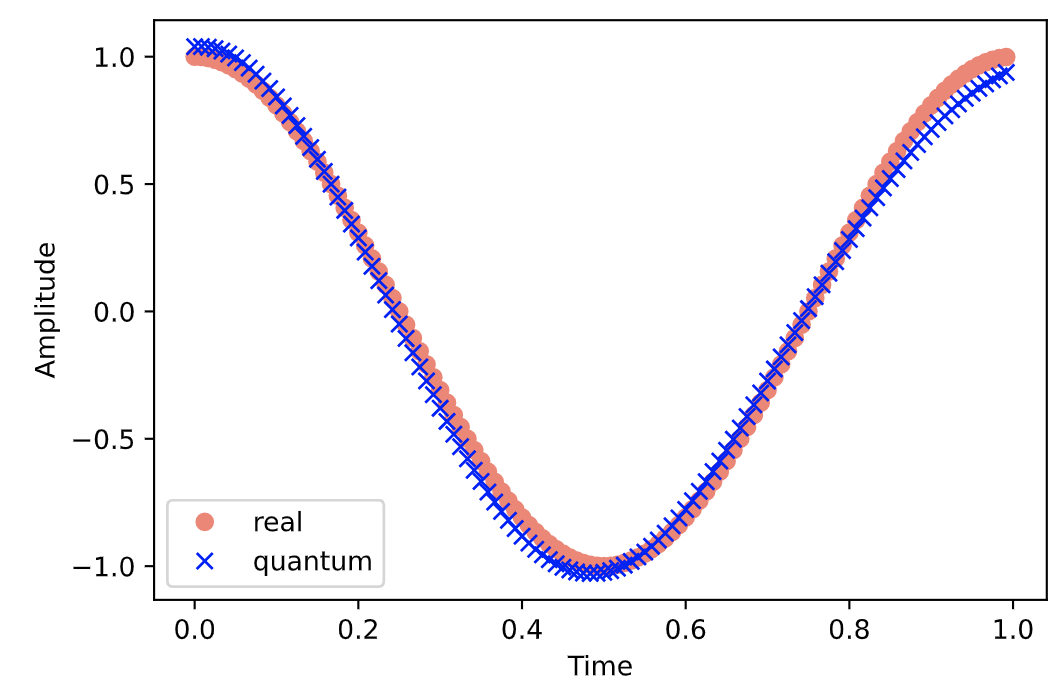}
		\includegraphics[width=0.2\textwidth]{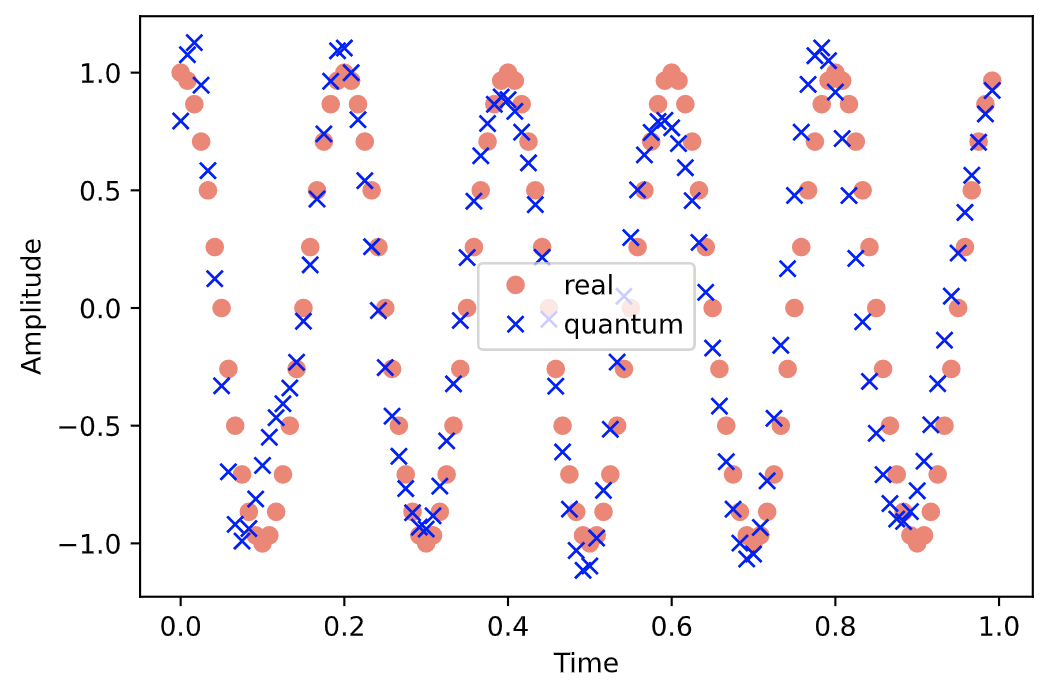}
        \includegraphics[width=0.2\textwidth]{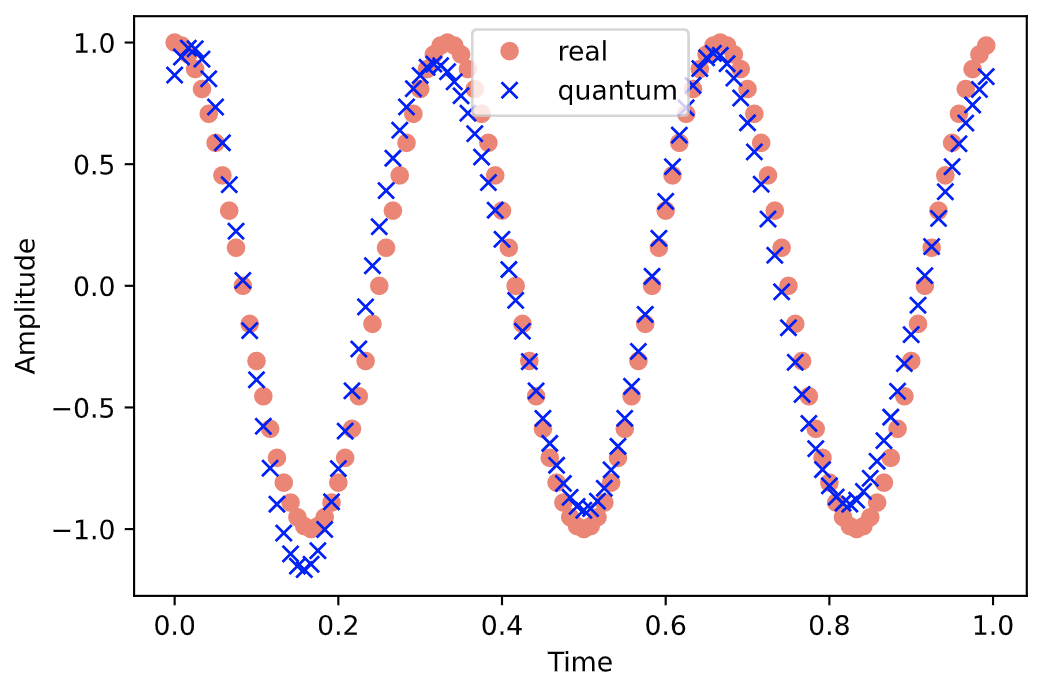} 
        \includegraphics[width=0.2\textwidth]{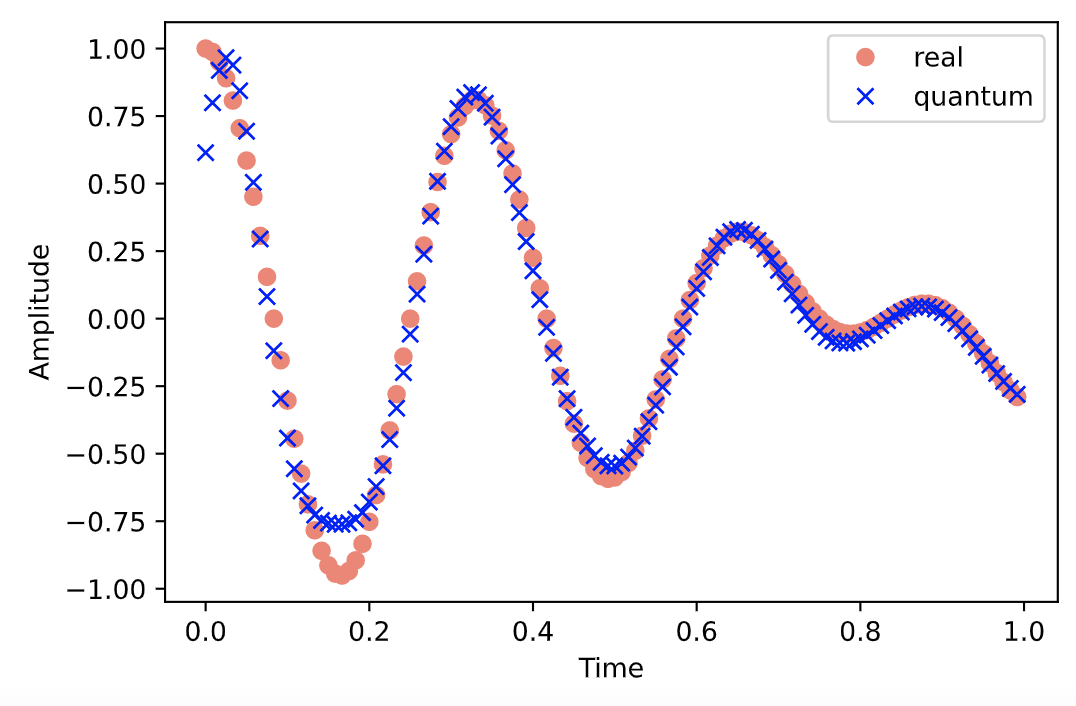}
		\caption{These figures show how the quantum model fits the sine and cosine input signal input. Several different frequencies have been tested; amplitude modulations have been tested to analyze how the model follows it. We realized the model limitation on the highest-frequency input signal.}
		\label{fig:resTrigoSine}
\end{figure}

This subsection presents the experiments' results using our trigonometric interpolation model. To test it, we operate as the input signal, the function \textit{sine, cosine, log, saw-tooth, and square}. Figures \eqref{fig:resTrigoSquare} to \eqref{fig:resTrigoSine} show the results from the trigonometric interpolation experiments. We define sample rate frequency to accomplish a Nyquist theorem \cite{landau1967sampling}. Next, we present the results. We have taken the opportunity to study the \textit{Gibbs phenomenon} \footnote{It describes the behavior of the Fourier series associated with a periodic piecewise-defined function in an unavoidable finite jump discontinuity.} in the square and saw-tooth signals. 

\begin{figure}[]
	\centering
        \includegraphics[width=0.4\textwidth]{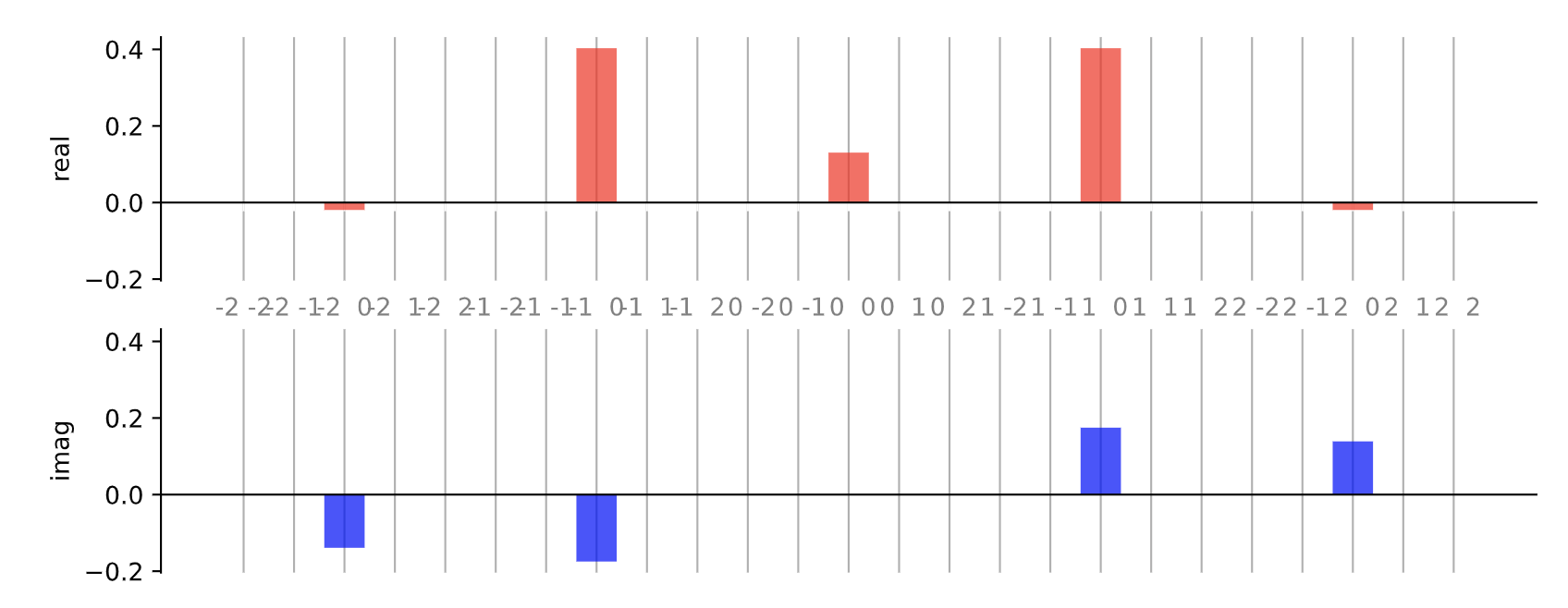}
		\caption{This graph illustrates the Fourier coefficients in the case of a $3Hz$ sinusoidal input signal, sampled above the Nyquist frequency at $120Hz$, and with $12$ repetitions, we can observe that the Fourier coefficients are potentially, in this case, rather real than imagined.
        Taking advantage of the \textit{Pennylane} \cite{Pennylane} tool, we validate our model provided by figure \eqref{fig:VQA_Fourier_Coeff}.}
		\label{fig:Fourier_coef}
\end{figure}

\begin{figure}[]
	\centering
        \includegraphics[width=0.4\textwidth]{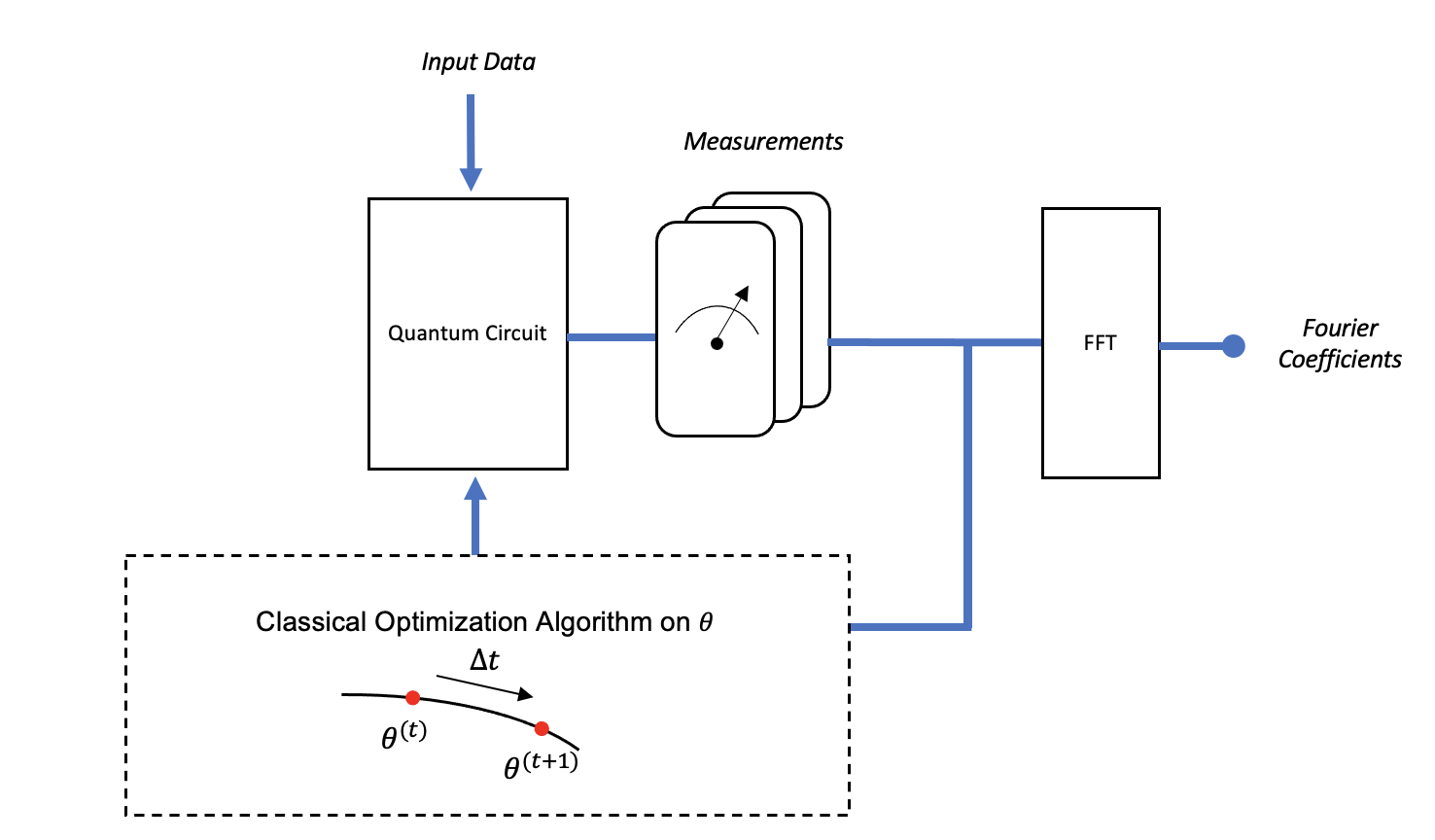}
		\caption{This diagram illustrates the steps to find the Fourier series coefficients using a variational model. The \textit{FFT} block samples the Fourier coefficients \eqref{fig:Fourier_coef} computing the first $2*K+1$ one of a $2*\pi$ periodic input signal, where $K \in N$.}
		\label{fig:VQA_Fourier_Coeff}
\end{figure}

\begin{figure}[]
	\centering
        \includegraphics[width=0.4\textwidth]{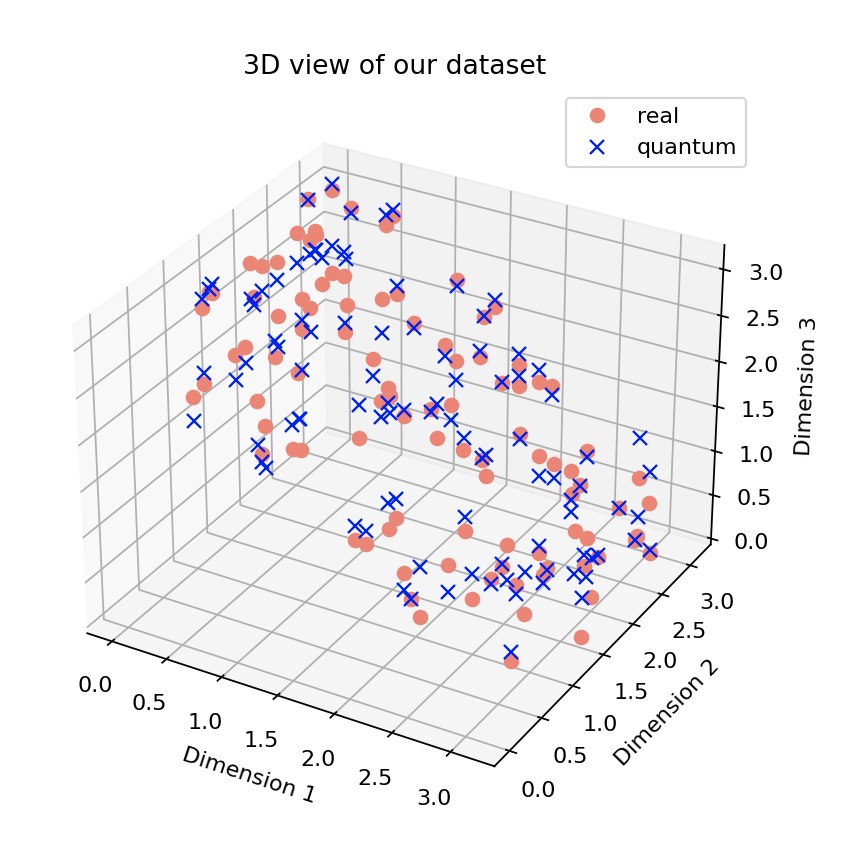}
		\caption{In this graph, we can analyze how we apply our model to data. We can observe how the model allows us to better interpolate the data from the dataset. One possible application is to find a continuous parameterized function from the discrete data. This approach can solve several real problems based on extensive classical data.}
		\label{fig:resTrigoDataset}
\end{figure}

\begin{figure}[]
	\centering
        \includegraphics[width=0.4\textwidth]{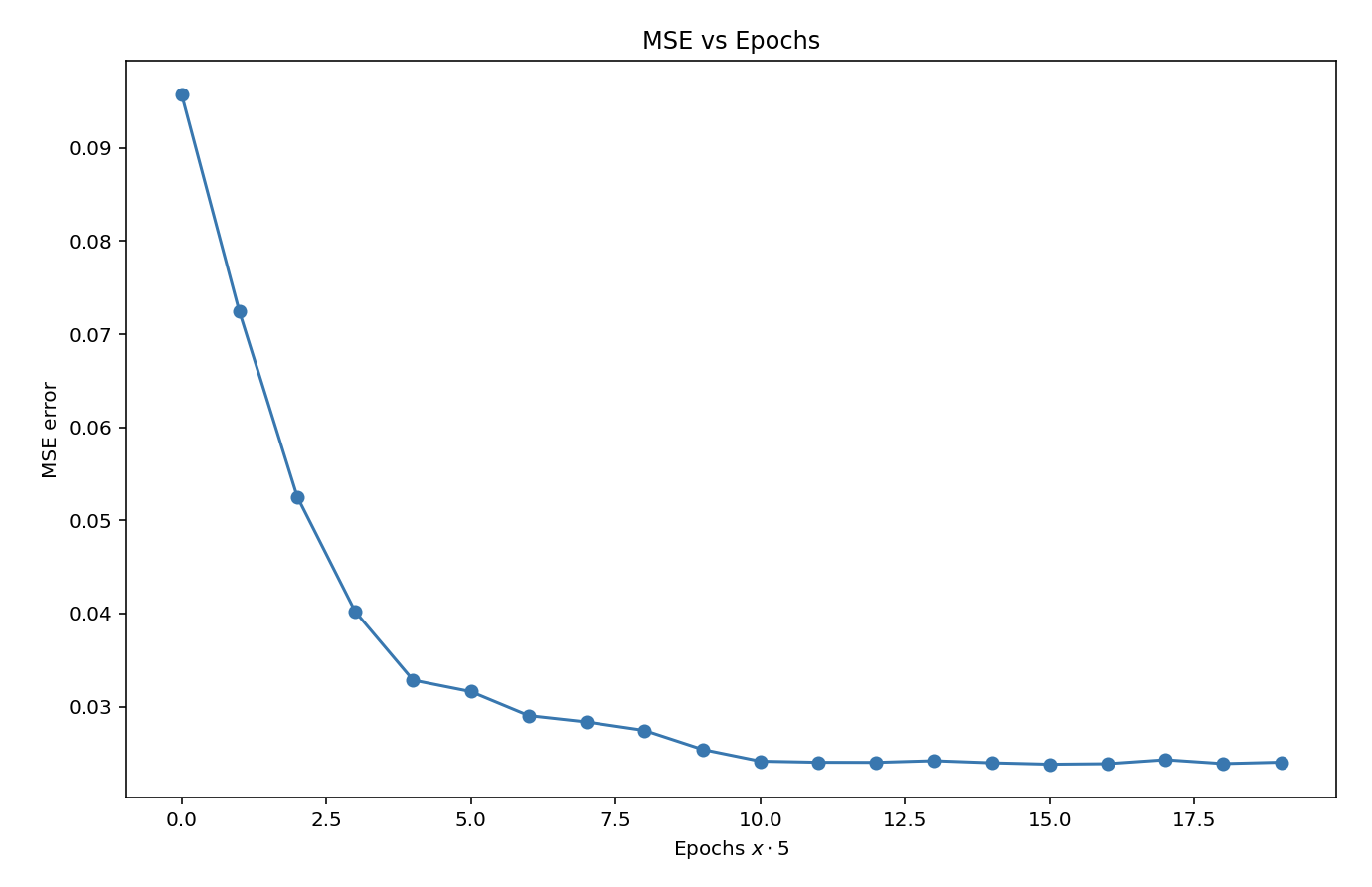}
		\caption{In this graph, we can see the evolution of our cost function. In this case, the equation described in \eqref{eq:err_qml}.}
		\label{fig:resTrigoDataset_err}
\end{figure}
\begin{figure}[]
	\centering
        \includegraphics[width=0.41\textwidth]{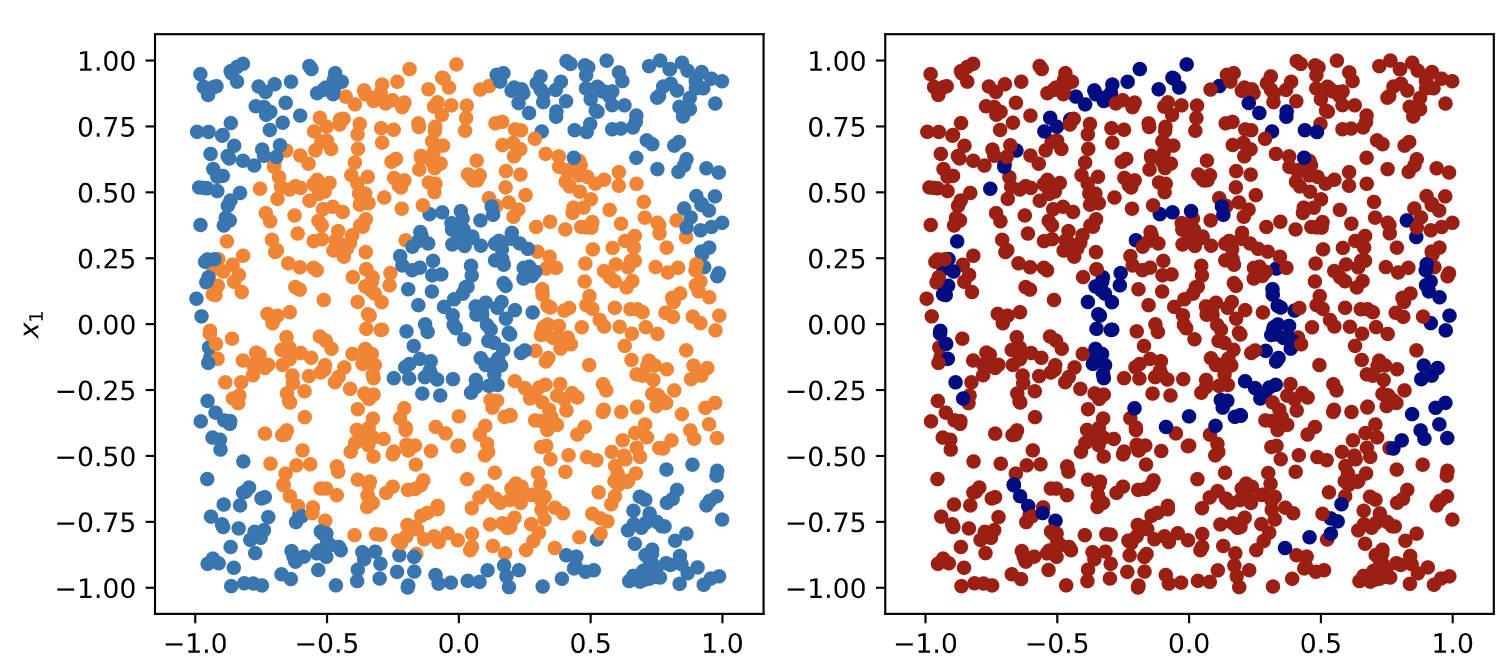}
        \includegraphics[width=0.41\textwidth]{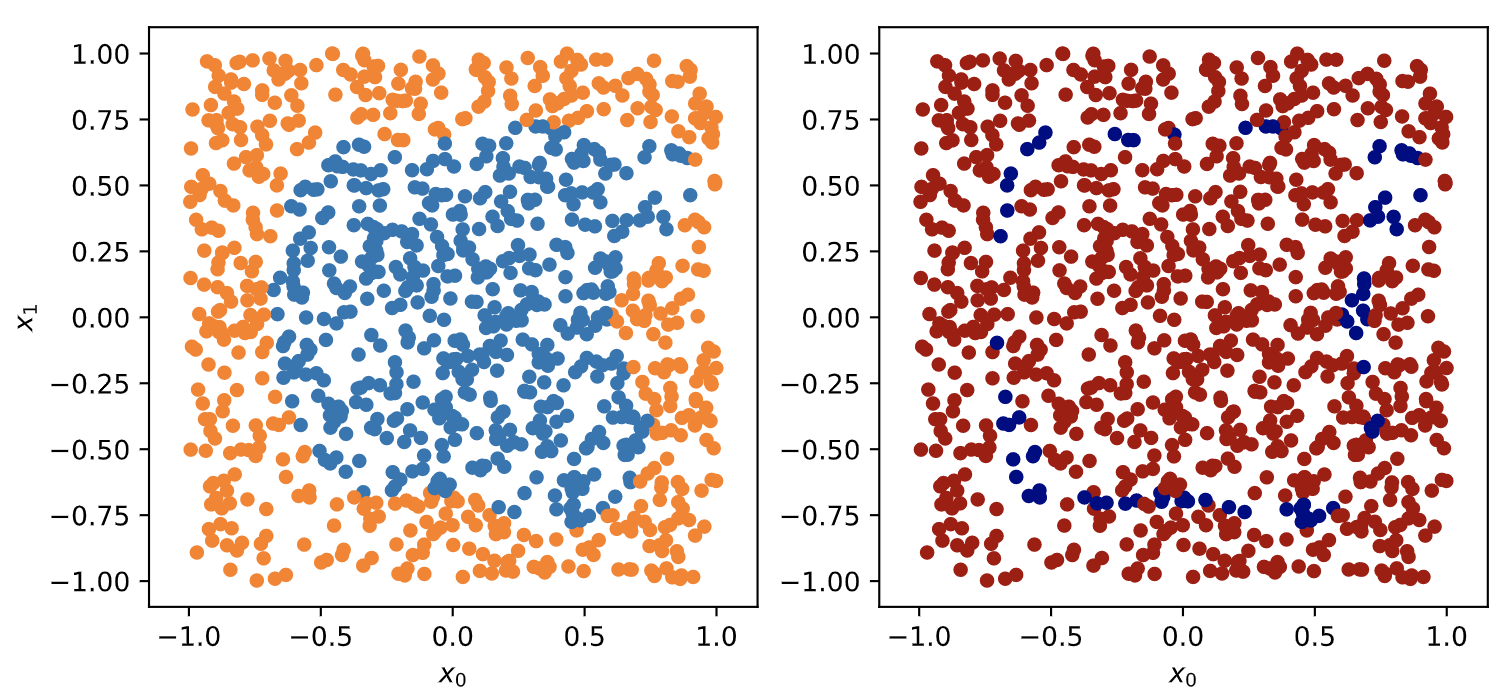}
		\caption{These figures are parts of the experiments we did in the case of the model as a classifier. In the first two graphs, we can observe that the figures above represent two circles. The one on the left is the classifier's result, and our model's classification errors are on the right. In this case, we are talking about $92.5\%$ accuracy. In the second two graphs, where the figure is to be classified in a square of others, we see the same in this case, an accuracy of $92.2\%$.}
		\label{fig:resCircle}
\end{figure}

\subsubsection{CLASSIFIERS}
In this subsection, we present the results of our experiments with our model in the classification section. To test the classifier, we have generated four datasets (crown, circle, square, SWP \cite{atchadeadelomou2022quantum, atchade2020using, adelomou2020formulation}). Figures, \eqref{fig:resTrigoDataset_err}, \eqref{fig:resCircle} and table \eqref{tab:results_qNN_1qbut} to \eqref{tab:results_CBR_NN_SW_Full} present the results of the experiments. 

\begin{table}[]
\centering
\begin{tabular}{ |c|c|c|  }
 \hline
 \multicolumn{3}{|c|}{  Quantum neural classifier ($1$ qubit) } \\
 \hline
 \#Layers  & \#Parameters & Accuracy\\
 \hline
    1  & 3 & 52.60\\
   2  & 6 & 59.90\\
   3  & 9 & 62.60 \\
  4  & 12 & 66.61\\
   5 & 15 & 63.72\\
   6 & 18 & 68.85\\
   7 & 21 & 67.91\\
   8 & 24 & 67.20\\
   10 & 30 & 73.91\\ 
 \hline
\end{tabular}
\caption {We can observe from the table and from the outcomes that the proposed model can be used to perform a classifier. This is the table from studying the binary classifier with two qubits. In this study, we consider the circuit depth ($l$), the number of the qubits $n$, and the number of the parameterized gates $m$, the number of parameters calculated as $n \times $m$ \times l $.  Where $m$ is the summation of the parameterized gates of the \textit{feature map} stage and the \textit{variational quantum algorithm}.} 
\label{tab:results_qNN_1qbut}
\end{table}

\begin{table}[]
\centering
\begin{tabular}{ |c|c|c|  }
 \hline
 \multicolumn{3}{|c|}{  Quantum neural classifier ($2$ qubits) } \\
 \hline
 \#Layers  & \#Parameters & Accuracy\\
 \hline
    1  & 6 & 48.90\\
   2  & 12 & 70.30\\
   3  & 18 & 84.5 \\
  4  & 24 & 84.90\\
   5 & 30 & 87.50\\
   6 & 36 & 92.10\\
   7 & 42 & 92.60\\
   8 & 48 & 92.50\\
   10 & 60 & 92.20\\
 \hline
\end{tabular}
\caption{We can observe from the table and from the outcomes that the proposed model can be used to perform a classifier. This is the table from studying the binary classifier with two qubits. In this study, we consider the circuit depth ($l$), the number of the qubits $n$, and the number of the parameterized gates $m$, the number of parameters calculated as $n \times $m$ \times l  $ where $m$ is the summation of the parameterized gates of the \textit{feature map} stage and the \textit{variational quantum algorithm}.}
\label{tab:results_qNN_2qbuts}
\end{table}

\begin{table}[]
\centering
\begin{tabular}{ |c|c|c|c|  }
 \hline
 \multicolumn{4}{|c|}{The SWP leveraged the quantum neural classifier.} \\
 \hline
 \#Patients & \#SW  &\#Layers   & Accuracy\\
 \hline
 3   & 2 &   2   & 70.3\\
 4   & 3 &   2   & 63.2\\
 5   & 2 &  2   & 57.7 \\
 5   & 3 &  2   & 48.1\\
 5   & 4 &  2  & 47.2\\
 \hline
\end{tabular}
\caption{We are solving the Social Workers' Problem (SWP) \cite{atchade2020using, adelomou2020formulation} with a quantum neural network classifier. }
\label{tab:results_CBR_NN_SW_Full}
\end{table}

\section{Discussions}\label{sec:discussions}
We delve deeper into the impactful outcomes of our study, building on the theoretical legacies established by previous researchers such as \cite{du2020expressive} and \cite{Schuld_2021}. Three core objectives drive our investigation: firstly, to empirically substantiate the intrinsic capability of quantum models in discerning and assimilating periodic patterns within datasets; secondly, to illuminate the crucial role of Fourier series representation in time series analysis and signal processing, especially its critical application in trigonometric interpolation for Quantum Machine Learning (QML).
Additionally, our research forays into the domain of quantum circuit design, drawing functional parallels with neural networks and underscoring its practical viability in interpolation (Figures. \eqref{fig:resTrigoSquare}, \eqref{fig:resTrigoSaw}, \eqref{fig:resTrigoSine} and \eqref{fig:resTrigoDataset}) and classification tasks (Fig. \eqref{fig:resCircle}). This exploration validates the theoretical underpinnings of quantum models and showcases their real-world applicability and versatility.
A key accomplishment of our work lies in bridging the theoretical and practical realms of quantum computing. We introduce a groundbreaking methodology for calculating Fourier coefficients, meticulously adapted for specific classification and trigonometric interpolation scenarios. This approach effectively narrows the divide between abstract mathematical concepts and their practical computational implementations in QML, harmonizing theoretical depth with pragmatic utility.
Our efforts aim to establish new benchmarks in QML. By demonstrating the effectiveness and adaptability of our methodology, we provide robust answers to the complex challenges of quantum data processing. This advancement validates our theoretical models and places our approach at the vanguard of QML. Our contributions will be a foundational reference, guiding future research and applications in the evolving intersection of quantum computing and machine learning.
Moreover, the proposed model in our study introduces a transformative approach to machine learning, with a particular emphasis on hyperparameter tuning as discussed in \cite{consulpacareu2023quantum}. This model surpasses traditional limitations, facilitating a more effective and expansive exploration of hyperparameter spaces. Utilizing the strengths of quantum computing, it enables a quicker, more thorough search of potential configurations, offering distinct advantages over classical methods, especially in situations constrained by time or computational resources.
Quantum computing also holds the potential to revolutionize optimization challenges by diminishing the search space and boosting performance. Its ability to handle vast solution spaces and conduct intricate calculations more efficiently could significantly expedite the process. This not only hastens the identification of solutions but also increases the probability of discovering optimal or near-optimal results. The integration of this model in machine learning applications underscores its significant impact in advancing computational techniques and addressing complex computational problems.

Future work will apply the ideas developed here in an accurate model. Although intuition leads us to think that the quantum model is a Fourier series, basically because the parameterized gates define the ansatz are rotation gates, the demonstration of this intuition is not trivial. Calculating the Fourier coefficients of the quantum model is also very challenging. In this paper, we have relied on an approximation and classical with the \textit{fft} to calculate them (see figure \eqref{fig:VQA_Fourier_Coeff}).
We have confirmed that the proposed model is a partial Fourier series. The model can be used for trigonometric interpolation (see Figures \eqref{fig:resTrigoSquare} to \eqref{fig:resTrigoDataset}), as seen in the figure \eqref{fig:resTrigoDataset}. 
Also, with a few subtle changes, we have been able to make a binary classifier (see figure \eqref{fig:resTrigoDataset_err} to \eqref{fig:resCircle}). It is worth determining the sign for a binary classifier to know what class it is in. We can perform it with expectation value (\textit{qml.expval}) or with probabilities (\textit{qml.probs}). In the case of having a multiclass classifier, what can be done is to generalize the binary classifier or work with the probabilities so that, knowing the highest probability of the measured outcome, the class will be known. Nevertheless, it is costly to go from binary classifier to multiclass; one must run $(n+1)n/2$ quantum models.

We have not made a detailed study of the algorithm's computational complexity; however, in case of a volume of data that presents a trigonometric distribution, the proposed model will be able to speed up the hyperparameter optimization process. This model can help to find hyperparameters more optimally and can help to have a continuous function model from discrete function models, thus offering more valuable solutions. 

\begin{figure}[b]
	\centering
        \includegraphics[width=0.2\textwidth]{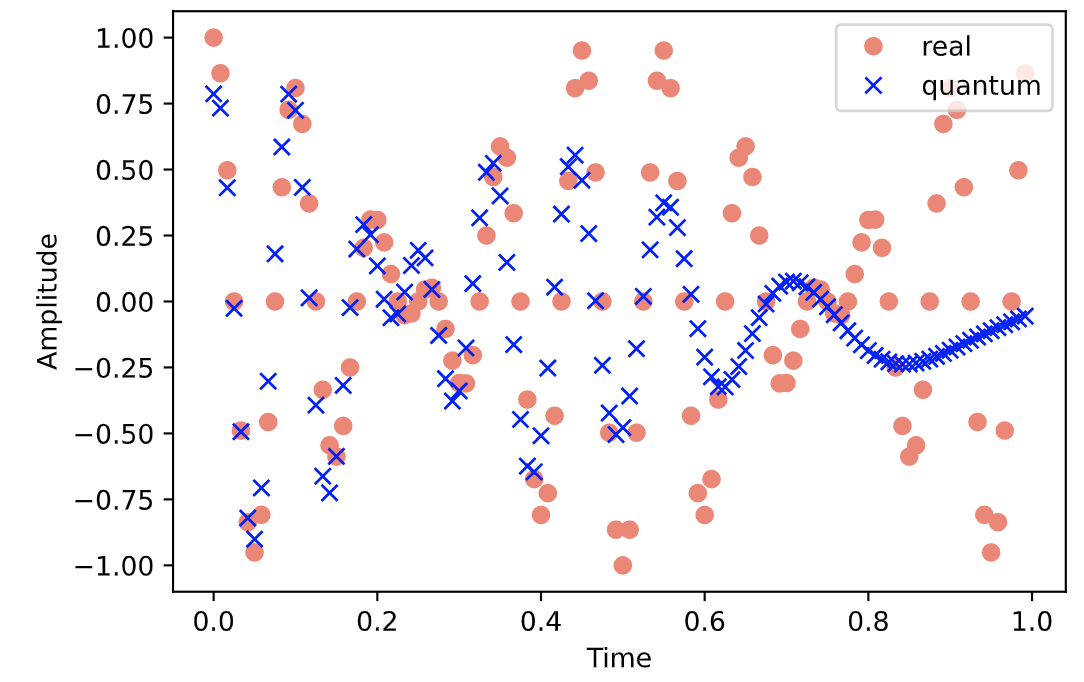}
        \includegraphics[width=0.2\textwidth]{images/Trigo/SineModula_10Hz_1Hz_30.png}
        \includegraphics[width=0.2\textwidth]{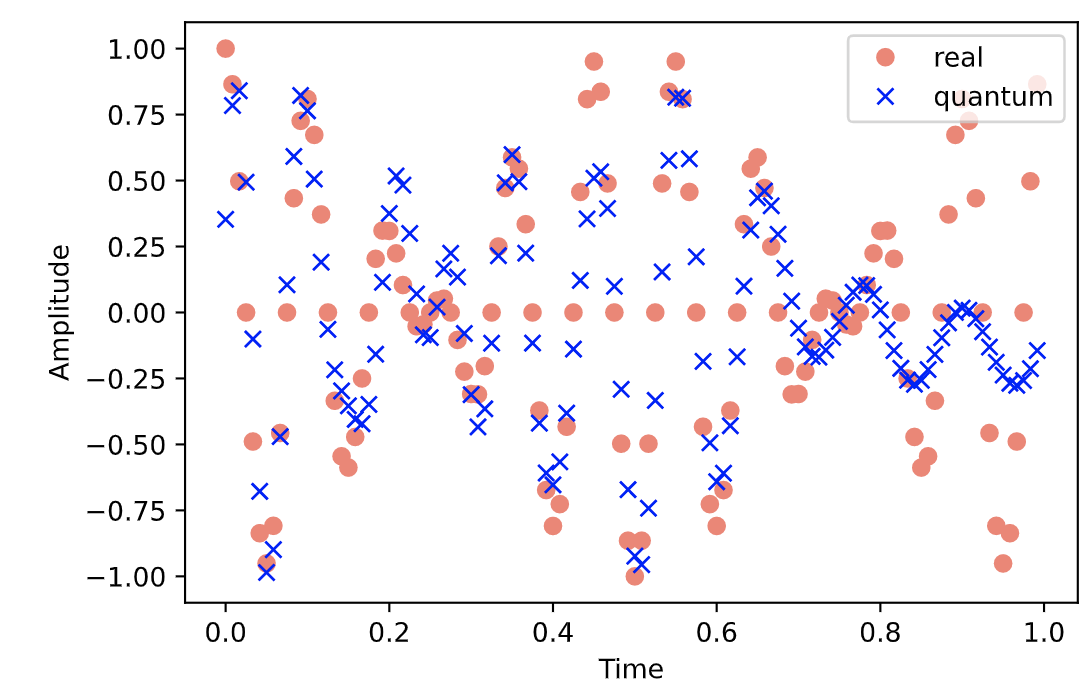}
        \includegraphics[width=0.2\textwidth]{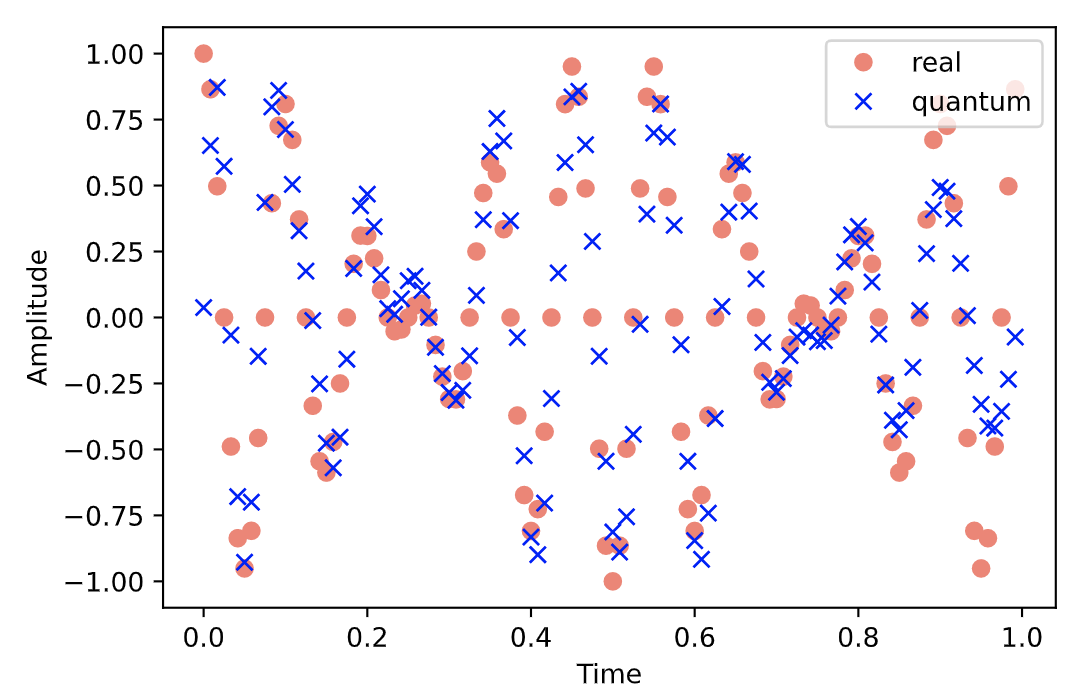}
		\caption{This graph presents the behavior experiment due to the limitations of high frequencies. The tail of the signal is seen to decay, and we must considerably increase the number of repetitions (layers) to solve it. In this graph, we multiply two quite different frequency sines. This is the result of simulating an amplitude modulation.}
		\label{fig:limitation}
\end{figure}

Tables \eqref{tab:results_qNN_1qbut} and \eqref{tab:results_qNN_2qbuts} give us detailed information on the quality of the classifier that we have managed to have some classifiers with accuracy greater than 92\%. From figure \eqref{fig:limitation}, which represents the amplitude modulation,  we can analyze the limitation of the proposed model for the highest frequencies where we can observe how the tail of the signal is seen to decay, and during the experimentation, we decided to increase the number of repetitions (layers) to adapt the model to the input signal.

\section{Conclusion} \label{sec:conclusions}
In this paper, we have experimented with and demonstrated the weight of the Fourier series in quantum machine learning. It also analyzed its impact on interpolations that can be used for banking problems, airline companies, retail companies, hyperparameter search problems, etc. We have also analyzed its impact on binary and multiclass classifiers, Hamiltonian simulation, and signal processing applications. The results conclude that we can strongly consider the quantum computer a Fourier series. However, we have also detected some limitations in better adapting the model to the highest-frequency input signals. Also, it remains a challenge to finish determining with greater precision the Fourier coefficients. The way we have found to determine the Fourier coefficients from the VQA model is given in figure \eqref{fig:VQA_Fourier_Coeff}.
As a line for the future, we want to apply this model to the real problems in the banking sector, mobility, etc. We envision two promising directions for future research, leveraging the insights gained from our study. The first path involves utilizing our findings in the realm of hyperparameter optimization. By employing the strategy of search state reduction, we can streamline the hyperparameter search process, potentially unlocking new efficiencies and capabilities in quantum computing models.
The second avenue for exploration stems from applying Fourier coefficients in evaluating the efficiency of quantum circuits. When data is encoded through Hamiltonian time evolution, it becomes possible to conceptualize the class of functions that quantum models can effectively learn as partial Fourier series. This approach highlights the natural propensity of quantum models to grasp periodic functions in data intuitively. Consequently, this necessitates the consideration of appropriate data re-scaling strategies, ensuring that the data aligns well within the period of the function class being modeled.
Furthermore, the role of classical pre-processing in enhancing the expressivity of smaller quantum models cannot be overstated. By creating additional features, classical pre-processing enriches the frequency spectrum of the data, thereby offering quantum models a richer tapestry of information to learn from and interpret. This can lead to a more nuanced and effective utilization of the quantum circuit’s capabilities.
Lastly, the freedom to adjust the entries of the observable has been a cornerstone in demonstrating the universality of quantum circuits. This approach to the quantum computing process enables the circuits to adapt and respond more effectively to the specific requirements of the data and the learning task at hand. As we look forward, these areas not only present fascinating opportunities for further research but also hold the potential to advance the field of Quantum Machine Learning significantly, bridging the gap between theoretical possibilities and practical applications.

\section*{Code} The reader can find the code in the following GitHub repository: \url{https://github.com/pifparfait/Fourier_Based_QML} to reproduce the figures and explore different settings of the proposed model.

\section*{Acknowledgements} PA wants to thank Guillermo Alonso Alonso De Linaje, Adrián Pérez-Salinas, Ameer Azzam, Daniel Casado Faulí, and Sergi Consul-Pacareu for the discussions.

\textbf{Compliance with Ethics Guidelines}\\
Funding: This research received no external funding. 

Institutional review: This article does not contain any studies with human or animal subjects.

Informed consent: Informed consent was obtained from all individual participants included in the study.

Data availability: Data sharing not applicable. No new data were created or analyzed in this study. Data sharing is not applicable to this article. 

\bibliography{main}
 
\end{document}